\documentclass[conference]{IEEEtran}
\usepackage{amssymb}
\usepackage{amsmath} 
\usepackage{amsfonts}
\usepackage{dsfont}
\usepackage{amsthm}
\usepackage{epsfig}
\usepackage{epstopdf}
\usepackage[linesnumbered,lined,boxed,commentsnumbered, ruled, algo2e]{algorithm2e}
\usepackage{mathrsfs}
\usepackage{algorithm}
\usepackage{algorithmic}
\usepackage{setspace}
\usepackage{bm}
\usepackage{tikz}
\usetikzlibrary{arrows}
\usepackage{subfigure}
\allowdisplaybreaks[4]
\usepackage{graphicx,booktabs,multirow}
\usepackage{epstopdf}
\usepackage{enumitem}
\usepackage{subfigure}
\usepackage{multirow}
\usepackage{tabularx} 
\usepackage{booktabs}
\usepackage{float}
\usepackage{color,xcolor}
\definecolor{colorhkust}{RGB}{20,43,140}
\definecolor{colortsinghua}{RGB}{116,52,129}
\definecolor{color1}{RGB}{128,0,0}

\newtheorem{remark}{Remark}

\begin{document}

\title{Federated Reinforcement Learning for Real-Time Electric Vehicle Charging and Discharging Control}
\author{
\IEEEauthorblockN{
$\text{Zixuan~Zhang}$\IEEEauthorrefmark{2}, 
$\text{Yuning~Jiang}$\IEEEauthorrefmark{4}, 
$\text{Yuanming~Shi}$\IEEEauthorrefmark{2},
$\text{Ye~Shi}$\IEEEauthorrefmark{2}, 
and
$\text{Wei~Chen}$\IEEEauthorrefmark{3}
}\\
\vspace{-0.4cm}
\IEEEauthorblockA{
\IEEEauthorrefmark{2}School of Information Science and Technology (SIST), ShanghaiTech University, China \\ 
\IEEEauthorrefmark{4}Automatic Control
Laboratory, École Polytechnique Fédérale de Lausanne (EPFL), Switzerland\\
\IEEEauthorrefmark{3}Department of Electronic Engineering, Tsinghua University, China \\ 
\setlength{}{Email: zxzhang@gmail.com, yuning.jiang@ieee.org, \{shiym,shiye\}@shanghaitech.edu.cn}, wchen@tsinghua.edu.cn}
}
\maketitle
\setlength\abovedisplayskip{2pt}
\setlength\belowdisplayskip{2pt}

\begin{abstract}
With the recent advances in mobile energy storage technologies, electric vehicles (EVs) have become a crucial part of smart grids. When EVs participate in the demand response program, the charging cost can be significantly reduced by taking full advantage of the real-time pricing signals. However, many stochastic factors exist in the dynamic environment, bringing significant challenges to design an optimal charging/discharging control strategy. This paper develops an optimal EV charging/discharging control strategy for different EV users under dynamic environments to maximize EV users' benefits. We first formulate this problem as a Markov decision process (MDP). Then we consider EV users with different behaviors as agents in different environments. Furthermore, a horizontal federated reinforcement learning (HFRL)-based method is proposed to fit various users' behaviors and dynamic environments. This approach can learn an optimal charging/discharging control strategy without sharing users' profiles. Simulation results illustrate that the proposed real-time EV charging/discharging control strategy can perform well among various stochastic factors.
\end{abstract}


\section{Introduction}
\begin{figure}[htbp!]
\centering
\includegraphics[width=0.9\linewidth]{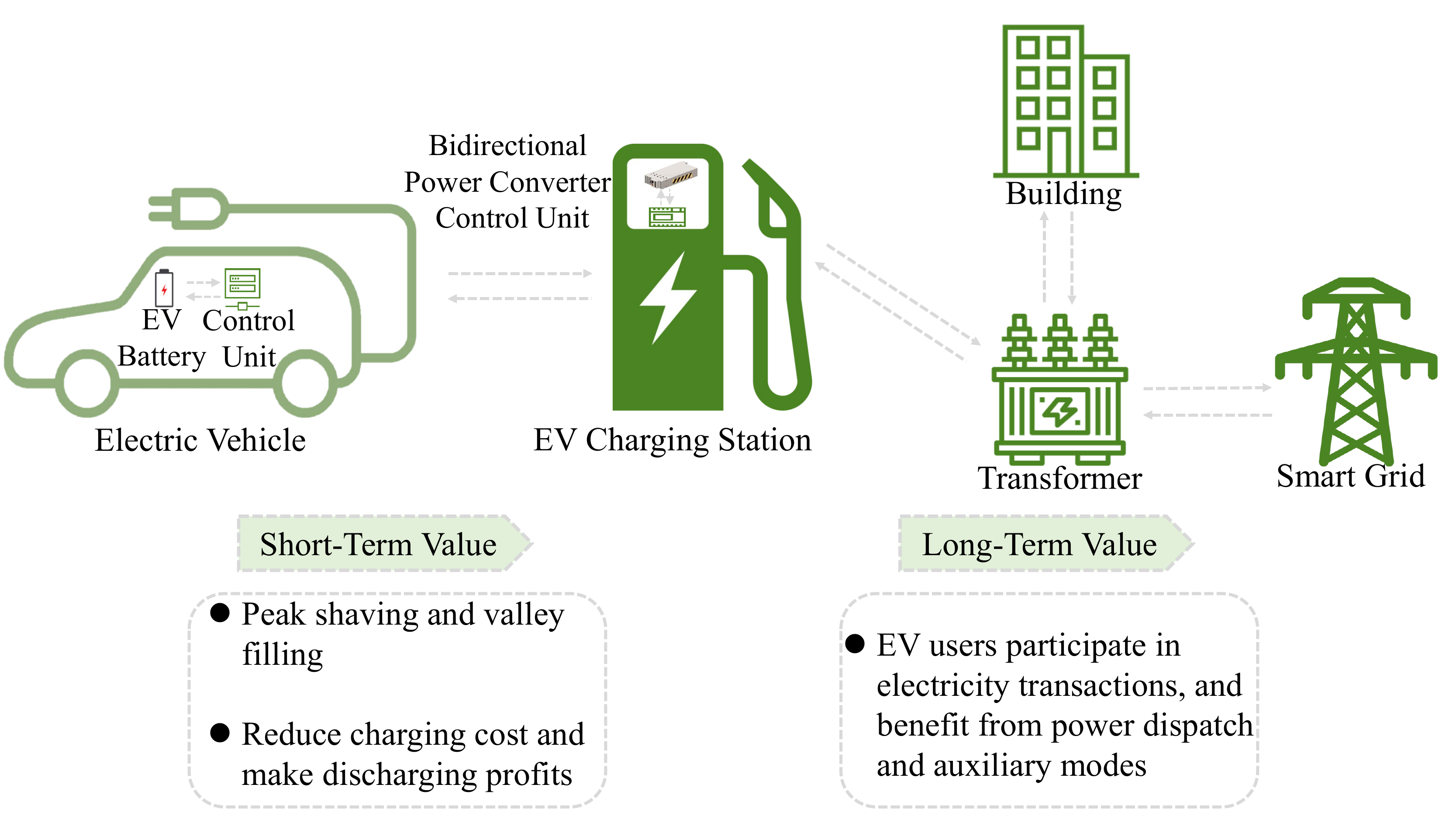}
\caption{V2G concept and its values}
\label{V2G}
\end{figure}

In the past decades, the advent of electric vehicles (EVs) has significantly mitigated air pollution and fossil energy depletion~\cite{chan2007state}. When EVs are connected to the power grid, they can serve in the discharging mode as vehicle-to-grid (V2G) devices or in charging mode as grid-to-vehicle (G2V) devices \cite{kempton2005vehicle}. As a new type of mobile and adjustable load, through switching their working modes alternatively, a fleet of EVs connected to the grid can work in the G2V mode at the valley time to reach valley filling and in the V2G mode at the peak time to achieve peak shaving \cite{gan2012optimal}. In addition, users' charging costs can be reduced by responding to the electricity price signals and changing the working pattern in time \cite{6102330}. The V2G concept and its benefits are shown in Fig. \ref{V2G}.

With the aim of maximizing users' benefits, the EV charging/discharging control strategy \cite{zhang2020cddpg} is always supposed to coordinate the charging/discharging action, including the charging/discharging decision and the charging/discharging rate. However, due to plentiful stochastic factors lying in the dynamic environment \cite{yan2021deep}, like time-varying electricity prices and uncertain behaviors from a variety of users, it is challenging to design an optimal charging/discharging control strategy considering many kinds of EV users.

Many day-ahead approaches, such as robust optimization (RO) \cite{ortega2014optimal} and stochastic optimization (SO) \cite{wu2017two} have been proposed to handle the price uncertainty. Although the methods above achieved great success in day-ahead charging/discharging control, it may be hard to depict the complex, real-time scenarios with more uncertain factors. Generally, the real-time charging/discharging control strategy considering uncertain electricity prices and users' demand satisfaction can be formulated as an optimization problem with a known state transition. Then it can be solved by model-based approaches like dynamic programming \cite{xu2016dynamic}, model predictive control (MPC) \cite{shi2018model}, model-based RL \cite{chics2016reinforcement}, etc. Nevertheless, it is tough to establish an accurate system model or estimate the state transition when considering 
various EV users' indeterminate charging/discharging behaviors.

As a technique that can directly learn the optimal policies without establishing or estimating the environment model \cite{chen2022reinforcement}, Model-free RL has been applied to numerous smart grid issues \cite{li2020real,mocanu2018line} and obtains good control performances. A model-free RL approach in \cite{9320523} can avoid grid congestion by coordinating EV charging/discharging. \cite{9716749} took the dynamic electricity price, non-EV residential load consumption, and drivers' behaviors into consideration to construct the dynamic environment. However, in these above papers, it is assumed that agents' state transitions follow the same distribution, i.e., the environments of different agents are IID. However, in actual scenarios, the situations faced by EV users may differ slightly, resulting in the non-IID environments and different state transitions.

As a novel type of distributed machine learning, federated learning (FL) \cite{letaief2021edge,yang2020federated,chen2020joint} has received considerable interest from academia and industry. FL allows the use of isolated data from multiple devices without violating the privacy protection policy \cite{shi2020communication,yang2020energy}, and it has been applied in many areas \cite{yang2022trustworthy,yang2022differentially,wang2021federated}. Recently, an emerging field called federated reinforcement learning (FRL) \cite{qi2021federated} combines the advantages of both FL and RL. It can not only provide agents with the experience to learn to make good decisions in unknown and dynamic environments but also train a global model collaboratively without sharing their own experiences. As a branch of FRL, horizontal federated reinforcement learning (HFRL) fits well for agents who are likely to isolate from each other but face similar decision-making issues and have fewer interactions \cite{zhang2021cooperative}.

This paper considers EV users with different behaviors as agents in different environments. Motivated by \cite{jin2022federated}, we aim to collaboratively learn a real-time EV charging/discharging control strategy that can perform uniformly well in different kinds of environments. We first formulate this problem as a Markov decision-making process (MDP), then a HFRL-based approach is proposed to deal with the dynamic charging/discharging environments and the users' various behaviors. In our approach, Soft Actor-Critic (SAC) algorithm \cite{haarnoja2018soft} can alleviate the sample-efficiency problem in RL as the local training method, and the FedAvg algorithm \cite{mcmahan2017communication} is utilized for global aggregation to help each agent quickly learn the optimal policy while considering privacy preservation. Moreover, our simulation results demonstrate that the proposed real-time EV charging/discharging control strategy can make a good trade-off between dynamic electricity prices and uncertain behaviors from different EV users.

\section{System Model and Problem Formulation}
This section first introduces a model to describe the dynamic changes of EVs' batteries. Then, the EV charging/discharging problem is formulated as a Markov decision process (MDP). Finally, we formulate the objective of the optimal charging/discharging control policy.

\subsection{EV Battery Model}
In this paper, we consider $N$ EVs equipped with the same batteries indexed by $i\in\{1,...,N\}$ and we assume that the charging infrastructures are the same for each EV. We define the times by $t_a^i$ at which EV $i$ arrives at the charging station and by $t_d^i$ at which it departs from the station. If the State of Charge (SoC) of EV $i$ at time $t$ and $t+1$ are denoted by $\mathrm{SoC}_t^i$ and $\mathrm{SoC}_{t+1}^i$, respectively, then the dynamic of $i$-th EV's battery between time instants $t$ to $t+1$ can be modeled as
\begin{equation}
\mathrm{SoC}_{t+1}^i=\left\{
\begin{array}{ll}
\mathrm{SoC}_t^i & t<t_a^i,\;t\geq t_d^i,\\[0.12cm]
\mathrm{SoC}_t^i+\eta \cdot a_t^i & t_a^i \leq t < t_d^i,\\
\end{array}
\right.
\label{dynamic_charge}
\end{equation}
where $a_t^i$ is the $i$-th EV's total charging/discharging rate during the time interval $[t, t+1)$. We assume the EV is under either V2G (charging mode, $a_t^i\geq0$) or G2V mode (discharging mode, $a_t^i\leq0$). Here, we also assume charging/discharging has the same efficiency $\eta \in (0,1]$ in this paper. Besides, $\mathrm{SoC}_t^i$ satisfies $\mathrm{SoC}_t^i\in[0, 1]$ for all $i$ and $t$, and the input $a_t^i$ is constrained by the charging infrastructure.

\subsection{MDP Formulation}
The EV charging/discharging control problem has the same form as the sequential decision-making problem, such that it can be regarded as a Markov decision process (MDP) with discrete steps. Let us consider EV users having different charging/discharging behaviors such that $N$ agents, respectively, interact with $N$ independent environments. The environments have different state transitions $\{\mathds{P}_i\}_{i=1}^N$ but the same state space $\mathcal{S}$, action space $\mathcal{A}$, reward function $\mathcal{R}$ and discount factor $\gamma$. Then the MDP of this problem can be denoted by $\mathcal{M}_i =\langle \mathcal{S}, \mathcal{A}, \mathds{P}_i, \mathcal{R}, \gamma \rangle$, for all $i=\{1,2,\cdots, N \}$.

\emph{1) State:} As the input of the charging/discharging control strategy, the environment state is used to generate a real-time charging/discharging action. For agent $i$,  the state $s_t^i\in\mathcal R^{n+6}$ at time $t$ includes the current and the past $n$ hours' electricity price $(\psi_{t-n}, \psi_{t-n+1}, \cdots, \psi_{t})\in\mathcal{R}^{n+1}$, the departure time $t_d^i$, the anxious time $t_{x}^i$, the current $\mathrm{SoC}_t^i$, the expected $\mathrm{SoC}_{x}^i$ at the anxious time and the departure time $\mathrm{SoC}_d^i$, that is
\begin{equation}\label{state}
s_t^i=\{\psi_{t-n}, \psi_{t-n+1}, \ldots, \psi_{t}, t_d^i,t_{x}^i, \mathrm{SoC}_t^i, \mathrm{SoC}_{x}^i, \mathrm{SoC}_d^i\}.
\end{equation}

\emph{2) Action:} The action $a_t^i$ denotes the charging/discharging rate of EV $i$'s battery during the state transition step $[t, t+1)$ with the given state $s_t^i$. Due to the limitation of charging infrastructures, the action is restricted as follows
\begin{equation}\label{action}
\underline{a}\leq a_t^i \leq \overline{a},
\end{equation}
where $\underline{a}$ and $\overline{a}$ are the minimum and maximum rate for this EV charging/discharging problem.

\emph{3) Reward:} The reward represents the immediate system feedback after the state $s_t^i$ changes to $s_{t+1}^i$ with $a_t^i$. We proposed a reward settlement scheme integrating EV's demand response factor into this reward function. A mathematical model in \cite{alsabbagh2020distributed} is used to quantify the effect of anxiety on SoC, and the anxious time $t_{x}^i$ is defined in this model. The model can be denoted as
\begin{equation}\label{anxiety}
\mathrm{SoC}_{x}^i=\frac{d_1^i\bigl(e^{-d_2^i(t-t_a^i)/(t_d^i-t_a^i)}-1\bigl)}{e^{-d_2^i}-1},
\end{equation}
where $t$ satisfies $t \in [t_{x}^i, t_d^i)$. It can map the user's anxiety to the expected SoC properly. $d_1^i \in [0, 1]$ and $d_2^i \in (-\infty, 0) \cup (0, \infty)$ are both shape parameters of the SoC curve. A larger $d_1^i$ leads to a higher SoC at $t_d^i$ and a larger $d_2^i$ determines a higher SoC during the charging/discharging duration. 

We assume that the price of selling and the purchasing electricity are the same, the reward can be defined as $r_t(s_t^i,a_t^i):=$
\begin{equation}\notag 
\begin{cases}
-\sigma_p \cdot \psi_{t}\cdot a_t^i & t_{a}^i\leq t < t_{x}^i,\\
-\sigma_p \cdot \psi_{t}\cdot a_t^i-\sigma_x\cdot \max(\mathrm{SoC}_{x}^i-\mathrm{SoC}_t^i, 0) &  t_{x}^i\leq t < t_d^i,\\
-\sigma_d \cdot \max(\mathrm{SoC}_d^i-\mathrm{SoC}_t^i, 0)  & t = t_d^i.\\
\end{cases}
\end{equation}
Here, the factors $\sigma_p$, $\sigma_x$, and $\sigma_d$ depict the user's sensitivity to price, anxiety, and demand response, respectively. When $t_a^i\leq t < t_{x}^i$, the reward fully considers the influence of electricity price fluctuations to EV charging/discharging decision. Then the EV user's anxiety is taken into consideration during $t_{x}^i\leq t < t_d^i$. As the EV leaves the charging station, i.e., $t=t_d^i$, the reward calculates the SoC gap between expected and current SoC to meet the EV user's demand as well as possible.

\emph{4) State Transition:} In practical charging/discharging scenarios, some personal factors like travel plans and  anxieties to price and SoC vary from person to person, i.e., their charging/discharging behaviors are different. To tackle this complicated situation, EV users with various charging/discharging behaviors can be regarded as agents located in different environments $\{\Omega_i\}_{i=1}^N$ with different state transitions $\{\mathds{P}_i \}_{i=1}^{N}$. Notice that establishing an accurate model considering dynamic electricity prices and users' charging/discharging behaviors is intractable, thus a model-free RL based algorithm is applied in the next section. It can learn a good control policy that does not rely on the certain model of the system.

\subsection{Problem Formulation}
Considering the various user charging/discharging behaviors mentioned before, we aim to learn an optimal charging/discharging control strategy $\pi^\star$ optimizing
\begin{equation}
\begin{aligned}
\max_{\pi}\biggl(\frac{1}{N}\sum_{i=1}^{N}\mathbb{E}&\Bigr[ \sum_{t=0}^{T}\gamma^t \cdot r_t(s_t^i,a_t^i) \\
&\bigg|a_t^i\sim \pi(\cdot|s_t^i), s_{t+1}^i \sim \mathds{P}_i(\cdot|s_t^i ,a_t^i)\Bigr]  \biggl)
\end{aligned}
\label{fedrl}
\end{equation}
which can fit well in different dynamic environments. Here, $\gamma \in [0,1]$ is the discount factor that relates to the rewards in the time domain. To handle the continuous state and action space in~\eqref{fedrl}, one needs to parameterize the Q-function and policy. Moreover, in order to obtain sufficient information in various environments for further training, each agent would be set to interact with the environment following its respective state transition. In the next section, we leverage Soft Actor Critic (SAC) approach to deal with the model parameterization and propose a federated framework to utilize all local data while preserving privacy. 

\section{Proposed Approach}
In this section, we propose a Soft Actor-Critic (SAC) based Horizontal Federated Reinforcement Learning (HFRL) method to deal with Problem~\eqref{fedrl}. To this end, we first give an overview of a generic HFRL framework, which can be summarized into two phases.
\begin{enumerate}
    \item \textbf{Local Phase} each agent (local client) $i$ in parallel do
    \begin{enumerate}[leftmargin =10pt]
        \item interacts with environment $\Omega_i$ and obtains the observation $s_t^i$. Then, generates action $a_t^i$ based on local policy $\pi_i(\cdot |s_t^i)$ and save the current reward $r_t^i(s_t^i, a_t^i)$. After that, collect the observation of next state $s_{t+1}^i$;
        \item updates the model of Q-function and policy using the local data. Then, send both local models to the central sever. 
    \end{enumerate}
    
    \item \textbf{Global Phase} central server receives all local information and aggregates them via \texttt{QAvg} and \texttt{PAvg}. Then the aggregated information is allocated to local clients.
\end{enumerate}
Here, the operator \texttt{QAvg} and \texttt{PAvg} replies on how to parameterize the local Q-function and policy $\pi_i$. In this paper, we approximate local Q-function approximation using SAC method such that the parametric local Q-function and policy can be defined by $Q_{\theta_i}$ and $\pi_{\phi_i}$ with $\theta_i$ and $\phi_i$ the parameters of DNN of actor and critic, respectively. Compared to the other existing parameterization approaches, SAC trains the local models with an entropy regularization to balance the exploration and exploitation \cite{haarnoja2018soft}. 

Using SAC to train the local models has two main steps, \textit{policy evaluation} for critic and \textit{policy improvement} for actor.
\subsubsection{Policy Evaluation} 
We need learn the soft Q-function
\begin{equation}\label{soft_q}
Q(s_t^i, a_t^i)=r_t(s_t^i, a_t^i)+\gamma\cdot \mathbb{E}_{s_{t+1}^i\sim p_i}\Bigr[ V(s_{t+1}^i) \Bigr],
\end{equation}
where the soft state value function $V(\cdot)$ is denoted by
\begin{equation}\label{soft_state_value}
V(s_t^i)=\mathbb{E}_{a_{t}^i\sim \pi_i}\Bigr[Q(s_t^i, a_t^i)-\alpha_i \log\bigl(\pi_i(a_t^i|s_t^i)\bigl) \Bigr].
\end{equation}
Essentially, SAC introduces a DNN $Q_{\theta_i}(s_t^i,a_t^i)$ with parameters $\mathbf{\theta}_i= (\theta_1^i, \theta_2^i, \cdots, \theta_n^i)$ to approximate the soft Q-function. Then the parameters can be trained by minimizing the soft Bellman residual by utilizing the previous sampled state and action data stored in the replay buffer $\mathcal{D}_i$, that is
\begin{equation}\label{critic_MSE1}
\mathcal{J}_Q(\theta_i)=\mathbb{E}_{(s_t^i,a_t^i)\sim\mathcal D_i}\left[ \frac{1}{2}\Bigl(Q_{\theta_i}(s_t^i,a_t^i)-\overline{Q}_{\bar\theta_i}(s_t^i,a_t^i)\Bigl)^2 \right]
\end{equation}
with $\overline{Q}_{\bar \theta_i}(s_t^i,a_t^i)$ given by
\begin{equation}\label{soft_Q}
\overline{Q}_{\bar\theta_i}(s_t^i,a_t^i)=r_t(s_t^i,a_t^i)+\gamma\cdot \mathbb{E}_{s^i_{t+1}\sim p_i}\bigr[V_{\bar{\theta}_i}(s_{t+1}^i) \bigr].
\end{equation}
Here, the $V_{\bar{\theta}_i}$ in (\ref{soft_Q}) is a DNN which can be used to estimate the soft state value function in (\ref{soft_state_value}), and its parameters can be updated by a moving average method, i.e., $\bar{\theta_i} \leftarrow \zeta\cdot \theta_i+(1-\zeta)\cdot \bar{\theta}_i$ with $\zeta \in (0, 1)$. Then $\mathcal{J}_Q(\theta_i)$ is optimized by using stochastic value gradient
\begin{align}\label{q_update}
&\widehat{\nabla}_{\theta_i}\mathcal{J}_{Q}(\theta_i):=\\\notag
&\quad \quad\nabla_{\theta_i}Q_{\theta_i}(s_t^i,a_t^i)\bigl(Q_{\theta_i}(s_t^i,a_t^i)-r_t(s_t^i,a_t^i)-\gamma V_{\bar{\theta}_i}(s_{t+1}^i)\bigl).
\end{align}
\subsubsection{Policy Improvement} We also use $\pi_{\phi_i}(a_t^i|s_t^i)$ as the approximation of the policy function $\pi_i(a_t^i|s_t^i)$. Then the optimal policy can be improved by minimizing the expected Kullback-Leibler (KL) divergence, that is
\begin{equation}\notag\label
{kl_divergence}
\mathcal{J}_{\pi}(\phi_i)=\mathbb{E}_{s_t^i\sim \mathcal{D}_i}\biggr[\mathbb{E}_{a_t^i\sim \pi_{\phi_i}}\Bigr[ \alpha_i \log\pi_{\phi_i}(a_t^i|s_t^i)-Q_{\theta_i}(s_t^i,a_t^i)  \Bigr]\biggr].
\end{equation}
Here, the temperature parameter $\alpha_i$ can be adjusted by an automating entropy method, that is minimizing
\begin{equation}
\label{alpha_update}
\Phi(\alpha_i):=\mathbb{E}_{a_t^i\sim \pi_{\phi_i}}\left[-\alpha_i\log \pi_{\phi_i}(a_t^i|s_t^i)-\alpha_i\widehat{\mathcal{H}}_i(\pi_{\phi_i})\right]
\end{equation}
over $\alpha_i$, where $\widehat{\mathcal{H}}_i$ is the desired minimum expected target entropy.
As a result of the charging/discharging rate $a_t^i$ and the current SoC $\mathrm{SoC}_t^i$ are both continuous, the policy $\pi_{\phi_i}$ is set as a Gaussian distribution
\begin{equation}
\label{gaussian_distribution}
\pi_{\phi_i}(a_t^i|s_t^i)=\frac{1}{\sqrt{2\pi \sigma}}\exp \Bigl(-\frac{(a_t^i-\mu)^2}{2\sigma^2} \Bigl), 
\end{equation}
where $\mu$ and $\sigma$ are both the ouput of the policy network. Then the reparameterization trick is employed to generate the current charging/discharging rate to support backpropagation. Specifically, if the mean is $\mu=\mu_{\phi_i}(s_t^i)$ and the standard deviation is $\sigma=\sigma_{\phi_i}(s_t^i)$, the policy can be reparameterized by $f_{\phi_i}(\kappa_t^i; s_t^i)= \mu_{\phi_i}(s_t^i) + \kappa_t^i \cdot \sigma_{\phi_i}(s_t^i)$, where $\kappa_t^i \sim \mathcal{N}(0,1)$. Through sampling $\kappa_t^i$ from the specific distribution and then intergrating the output of policy network, the charging/discharging rate $a_t^i$ can be generated.    
Finally, we can employ policy gradient based approaches by using the following unbiased gradient approximation of $J_\pi(\phi_i)$
\begin{align}\label{actor_update}
&\widehat{\nabla}_{\phi_i}\mathcal{J}_{\pi}(\phi_i)=\nabla_{\phi_i}\alpha_i\log \phi_i(s_t^i,a_t^i)\\\notag
&+\bigl(\nabla_{a_t^i}\alpha_i\log \pi_{\phi_i}(a_t^i|s_t^i)- \nabla_{a_t^i}Q(s_t^i,a_t^i)\bigl)\nabla_{\phi_i}f_{\phi_i}(\kappa_t^i; s_t^i).
\end{align}

\begin{remark}
In practice, we use two parameterized soft Q-function $Q_{\theta_i^k}(s_t^i,a_t^i)$ and two parameterized soft state value function $V_{\bar{\theta}_i^k}(s_{t+1}^i)$, where $k=\{1,2\}$ to mitigate positive bias. For the critic network $Q_{\theta_i^k}$, the gradient is calculated as follows
\begin{equation}
\label{actual_critic_update}
\mathcal{J}_Q(\theta_i^k)=\mathbb{E}_{(s_t^i,a_t^i)\sim \mathcal{D}_i}\biggr[ \frac{1}{2}\Bigl(Q_{\theta_i^k}(s_t^i,a_t^i)-{Q}_{\min}\Bigl)^2 \biggr],
\end{equation}
where $Q_{\min}$ satisfies
\begin{equation}
\label{q_min}
Q_{\min}=r_t(s_t^i,a_t^i) +\gamma \min_{k=1,2} \mathbb{E}_{s^i_{t+1}\sim p_i}\bigr[V_{\bar{\theta}_i^k}(s_{t+1}^i) \bigr].
\end{equation}
Besides, a parameterized Gaussian policy network $\pi_{\phi_i}(a_t^i|s_t^i)$ is also adopted in order to generate the charging/discharging action. In this way, we can finish the local update process.
\end{remark}
\begin{algorithm}[htbp!]
\footnotesize
\caption{Horizontal Federated Reinforcement Learning-Based Approach for EV charging/discharging Control}
\begin{algorithmic}[1]
\STATE \textbf{Input}: $\phi_i$, $\theta_i^k$, $\bar{\theta}_k^i$
\STATE \textit{Local Initialization:}
\FOR{$i=1$ \KwTo $N$ \textbf{each agent in parallel}}
    \STATE Initialize the actor network weights $\phi_i$ randomly.
    \STATE Initialize the replay buffer $\mathcal{D}_i$.
    \FOR{$k=1$ \KwTo $2$}
        \STATE Initialize the critic network weights $\theta_i^k$ randomly.
        \STATE Initialize the target network weights $\bar{\theta}_k^i$ using $\theta_i^k$. 
   \ENDFOR
\ENDFOR
\STATE \textit{Federated Training Process:}
    \FOR {episode = 1 \KwTo 250}
        \STATE \textit{Parallelizable Phase I: local state transition}
        \FOR {$i=1$ \KwTo $N$ \textbf{each agent in parallel}}
            \STATE When $\text{episode} > 1$, after receive $\theta_g^k$, $k=1,2$ and $\phi_g$, update the local model by $\theta_i^k\leftarrow \theta_g^k$ and $\phi_i\leftarrow \phi_g$.
            \STATE Get action $a_t^i$ in terms of state $s_t^i$ using $\pi_{\phi_i}$.
            \STATE Execute $a_t^i$, obtain reward $r_t(s_t^i,a_t^i)$ and state $s_{t+1}^i$.
            \STATE Store the tuple $(s_t^i, a_t^i, r_t(s_t^i,a_t^i), s_{t+1}^i)$ into $\mathcal{D}_i$.
        \ENDFOR
        
        \STATE \textit{Parallelizable Phase II: local model update}
            \FOR {$i=1$ \KwTo $N$ \textbf{each agent in parallel}}
            \STATE Update $\alpha_i$ by minimizing $\Phi(\alpha_i)$ in (\ref{alpha_update}).
            \STATE Update $\pi_{\phi_i}$ using gradient in (\ref{actor_update}).
            \FOR{$k=1$ \KwTo 2}
                \STATE Update $Q_{\theta_i^k}$ using gradient in (\ref{q_update}).
                \STATE Update $V_{\bar{\theta}_i^k}$ using the moving average method.
            \STATE Upload the local critic models $\theta_i^k$ to the server.
            \ENDFOR
            \STATE Upload the local actor models $\phi_i$ to server.
            \ENDFOR
            
            \STATE \textit{Aggregative Phase III: global model update}
            \STATE The server waits until receives all local models, then makes model aggregation
            $\theta_g^k = \frac{1}{N}\sum_{i=1}^N \theta_i^k$, $k=1,2$ and $\phi_g = \frac{1}{N}\sum_{i=1}^N \phi_i$.
            
            \STATE The server broadcasts $\theta_g^k$, $k=1,2$ and $\phi_g$ to each agent.
    \ENDFOR
\end{algorithmic}
\label{algo1}
\end{algorithm}

The proposed HFRL-based approach is summarized in Algorithm \ref{algo1}. First, all the parameterized policy and value networks are initialized locally. Then during the \textit{federated training process}, each agent directly uses the initialized policy to generate actions at the first episode and otherwise, first updates the local model by using the global model received from the central server. Then, they can, in parallel, collect experience from the local environment through the state-action-reward-state cycle interactions and store the experience into replay buffers as the tuple form in the \textit{local state transition} phase. After that, in the \textit{local model update} phase, the minibatch of experience is sampled from replay buffers to update these models locally and then transmitted to updated local models to the central server. In the \textit{global model update} phase, all the local models are aggregated, and the resulting global models are then sent back to the agents. 
\begin{figure}[htbp!]
\centering
\includegraphics[width=0.9\linewidth]{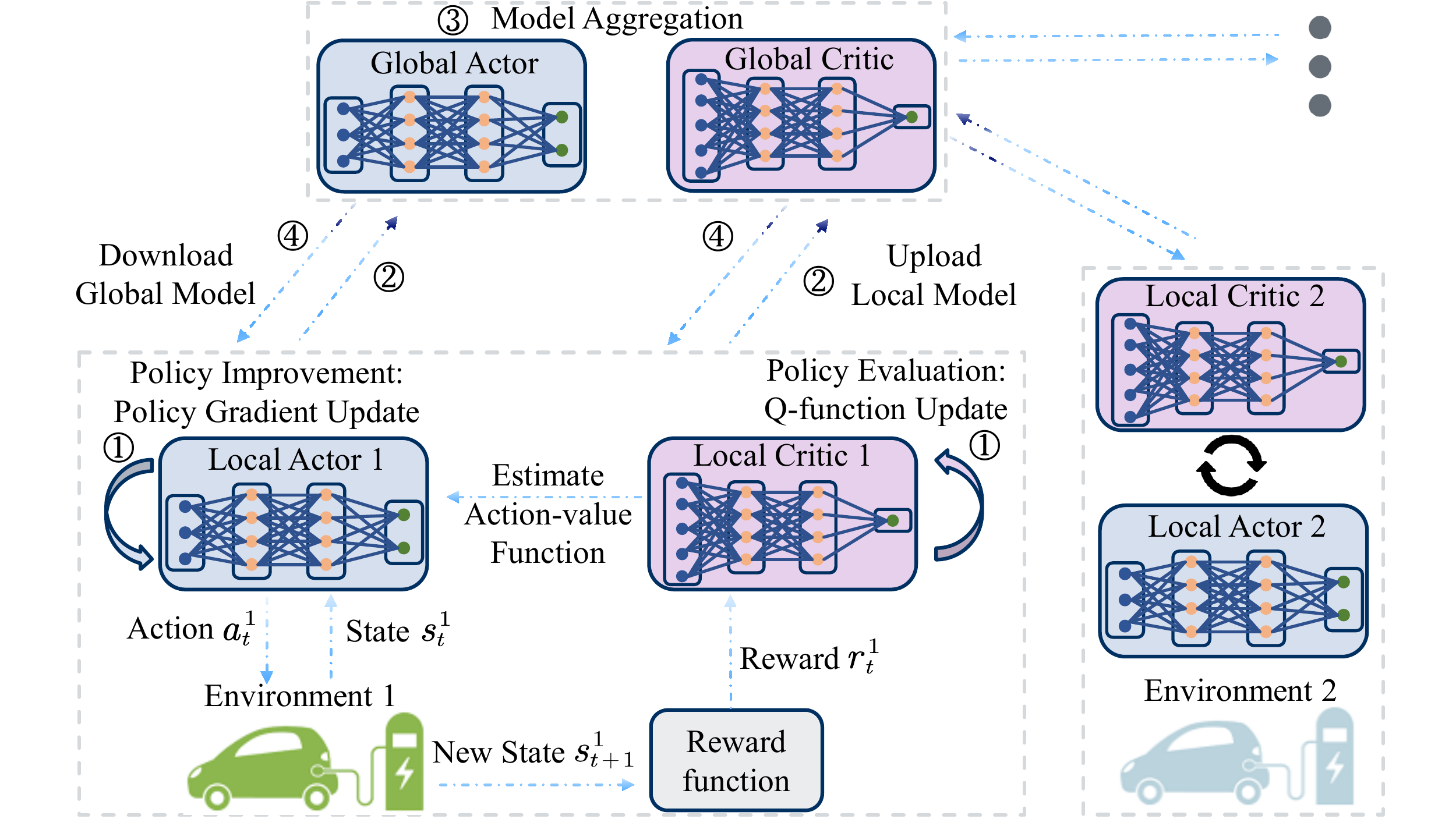}
\caption{A HFRL scheme for real-time EV charging/discharging control}
\label{frl_framework}
\end{figure}

Fig.~\ref{frl_framework} visualizes the proposed HFRL framework for EV charging/discharging control policy. Each EV can obtain its current charging/discharging rate according to the observed state by the local actor. After collecting the charging/discharging profiles for a period of time, the performance of actor and critic models can be improved by the policy evaluation and policy improvement procedure. Then through the local model upload, model aggregation, and model download procedure, one training episode is over. After many training episodes, the real-time charging/discharging control strategy considering different user behaviors is well trained. Since this method does not share users' charging/discharging profiles with the central server, it can significantly protect the privacy of EV users. As the standard federated framework, the proposed algorithm can learn a collaborative global model without collecting the local experience such that it preserve the EV users' privacy. 

\section{Results and Simulation}
In this section, we introduce some basic training settings for the HFRL-based approach illustrated in section \uppercase\expandafter{\romannumeral3} and show the training results. Then we verify the performance of the collaborative training model by one week's trip simulation.

\subsection{Training Settings and Results}
We use the real-world regional hourly electricity price data as a part of the state to extract the price trend. The price data from Jan 1st, 2017, to June 19th, 2017, is downloaded from the California ISO \cite{oedi_4033}. We select the first 20 days' price data each month for training and the remaining for evaluation. Moreover, three EV users with different charging/discharging behaviors are set up to interact with environments. For these users, the initial SoC by arrival time $t_a^i$ is sampled uniformly from $[0, 0.95]$, and the maximum charging/discharging rate during each state transition step accounts for 20\% of the EV capacity. The efficiency factor $\eta$ is equal to 0.98. The shape parameter $d_2^i$ in (\ref{anxiety}) is sampled from $\mathcal{N}(9,1^2)$ and the bound is $[6, 12]$. In addition, the sensitivity factors in the reward $r_t(s_t^i,a_t^i)$ are selected as $\sigma_p=8$, $\sigma_x=15$ and $\sigma_d=35$.
\begin{figure}[htbp!]
\centering
\subfigure[Home arrival]{
\includegraphics[width=0.29\linewidth]{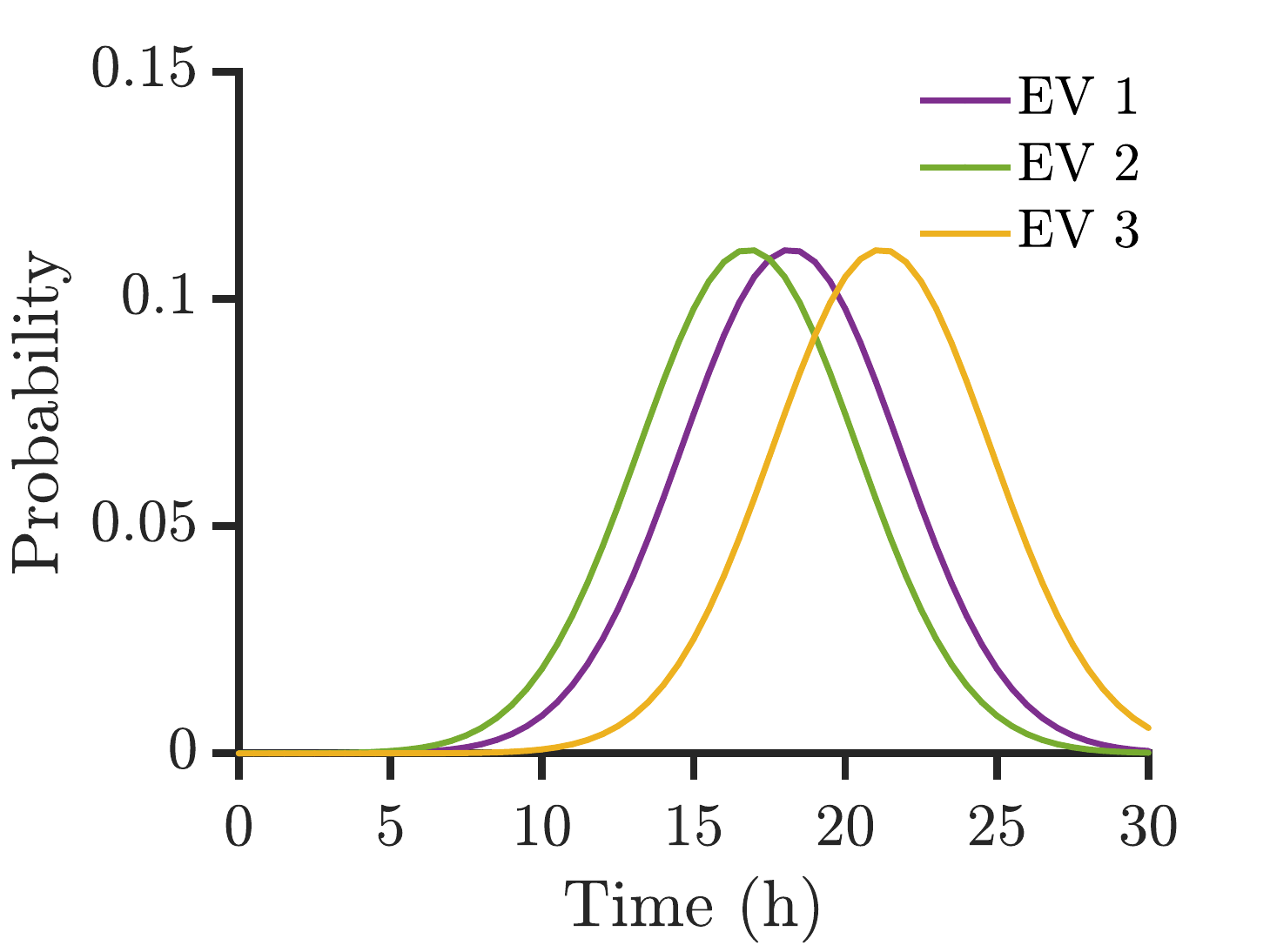}
}
\subfigure[Office arrival]{
\includegraphics[width=0.29\linewidth]{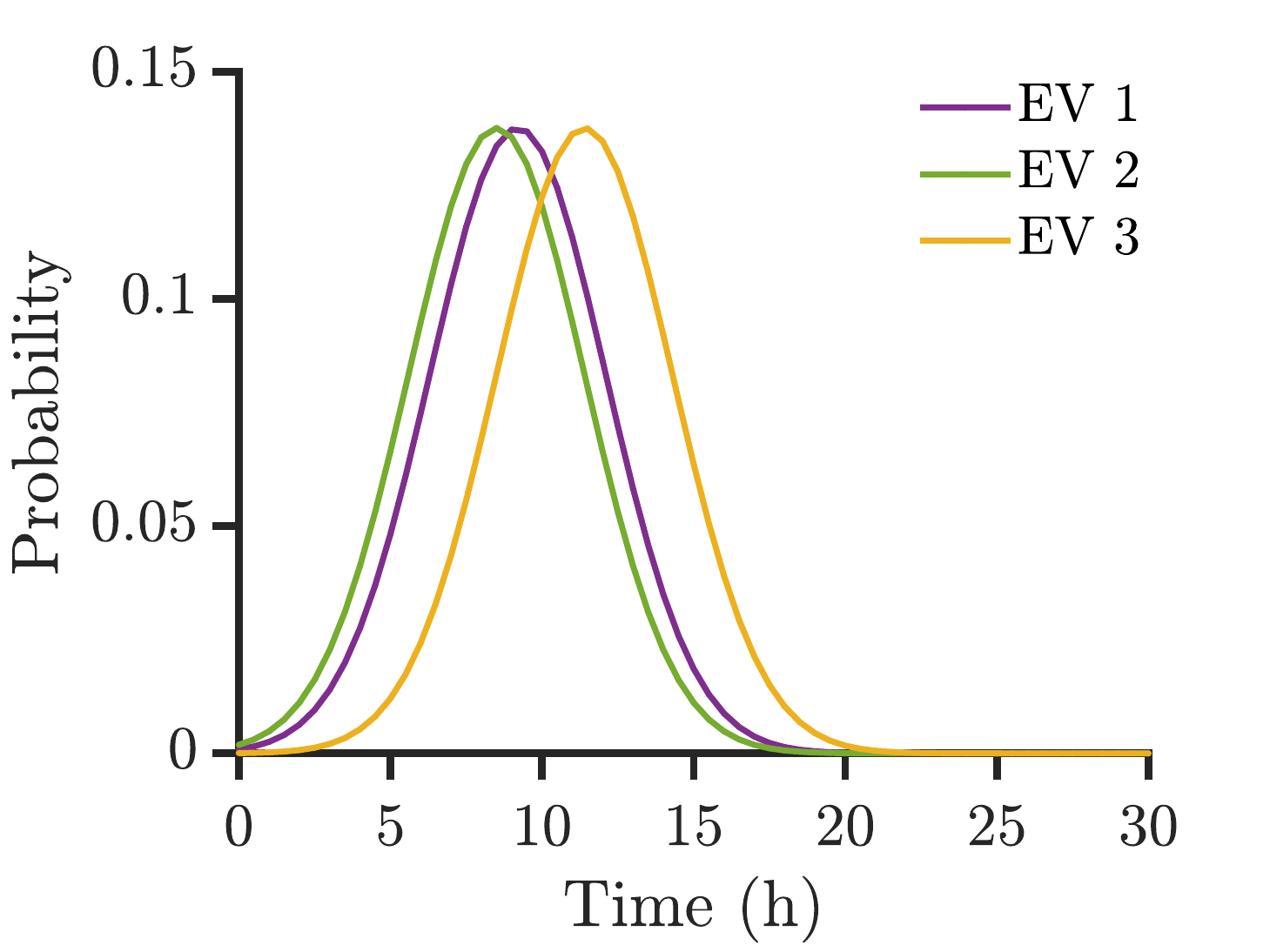}
}
\subfigure[Public arrival]{
\includegraphics[width=0.29\linewidth]{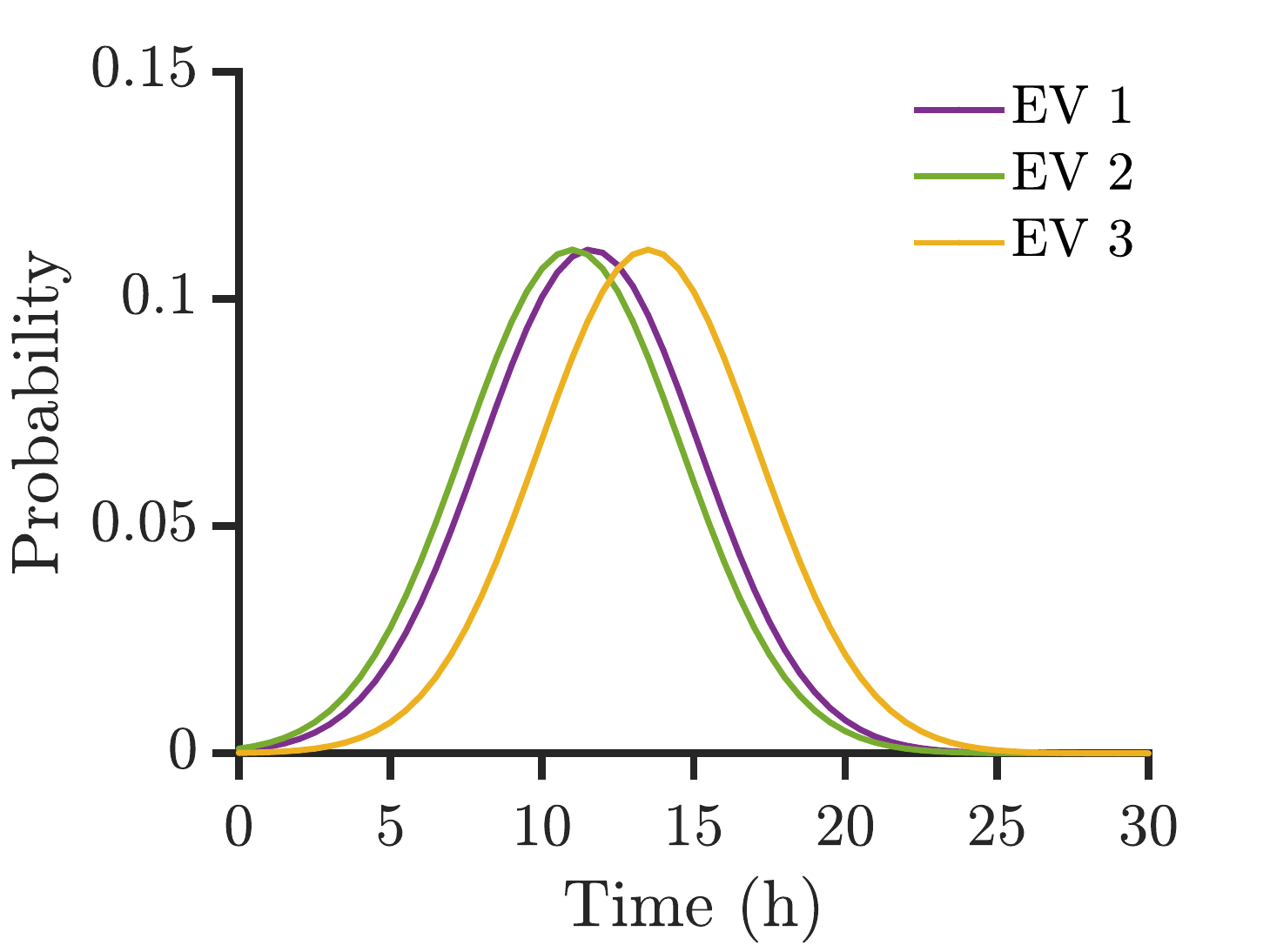}
}
\subfigure[Home departure]{
\includegraphics[width=0.29\linewidth]{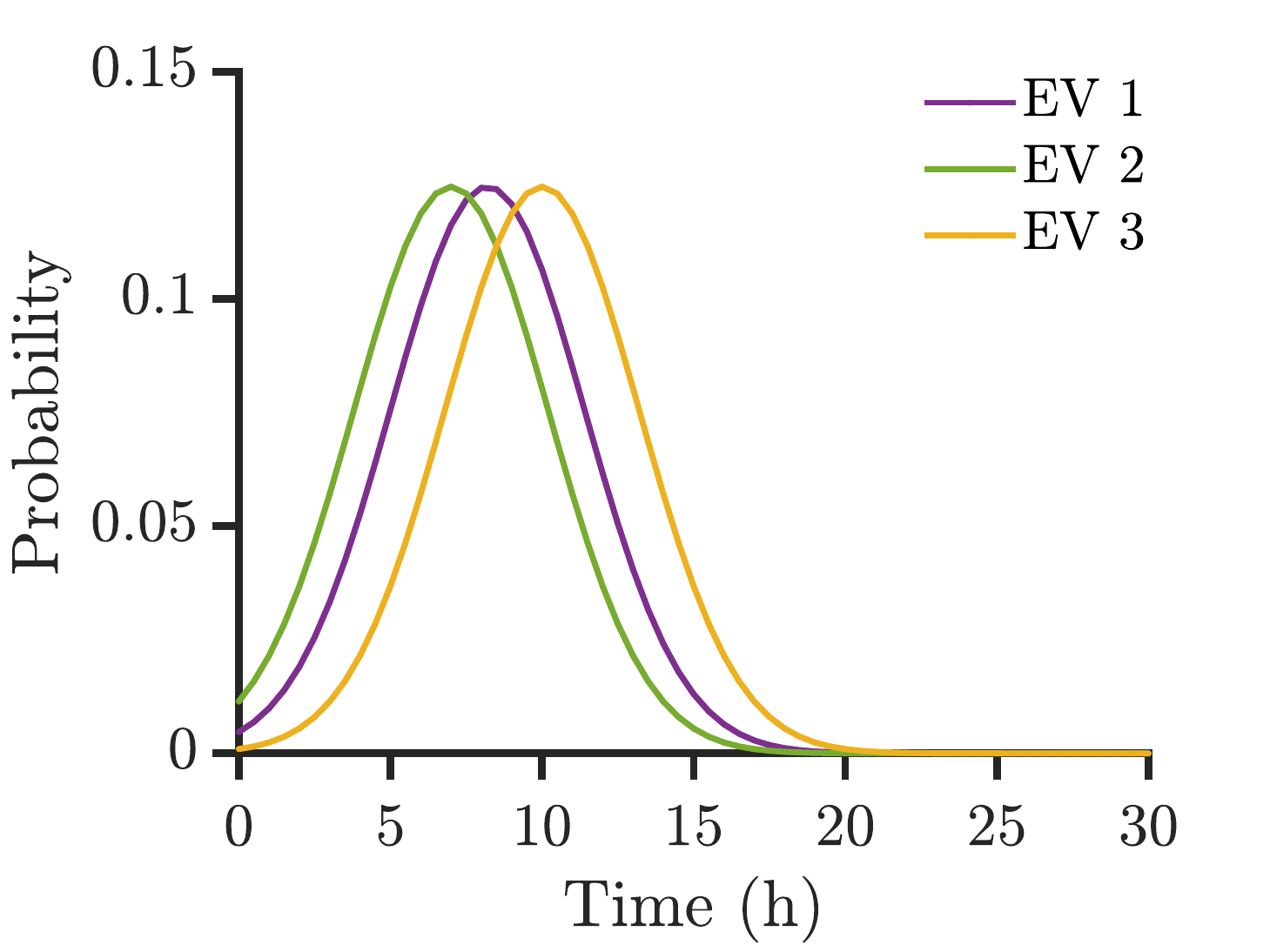}
}
\subfigure[Office departure]{
\includegraphics[width=0.29\linewidth]{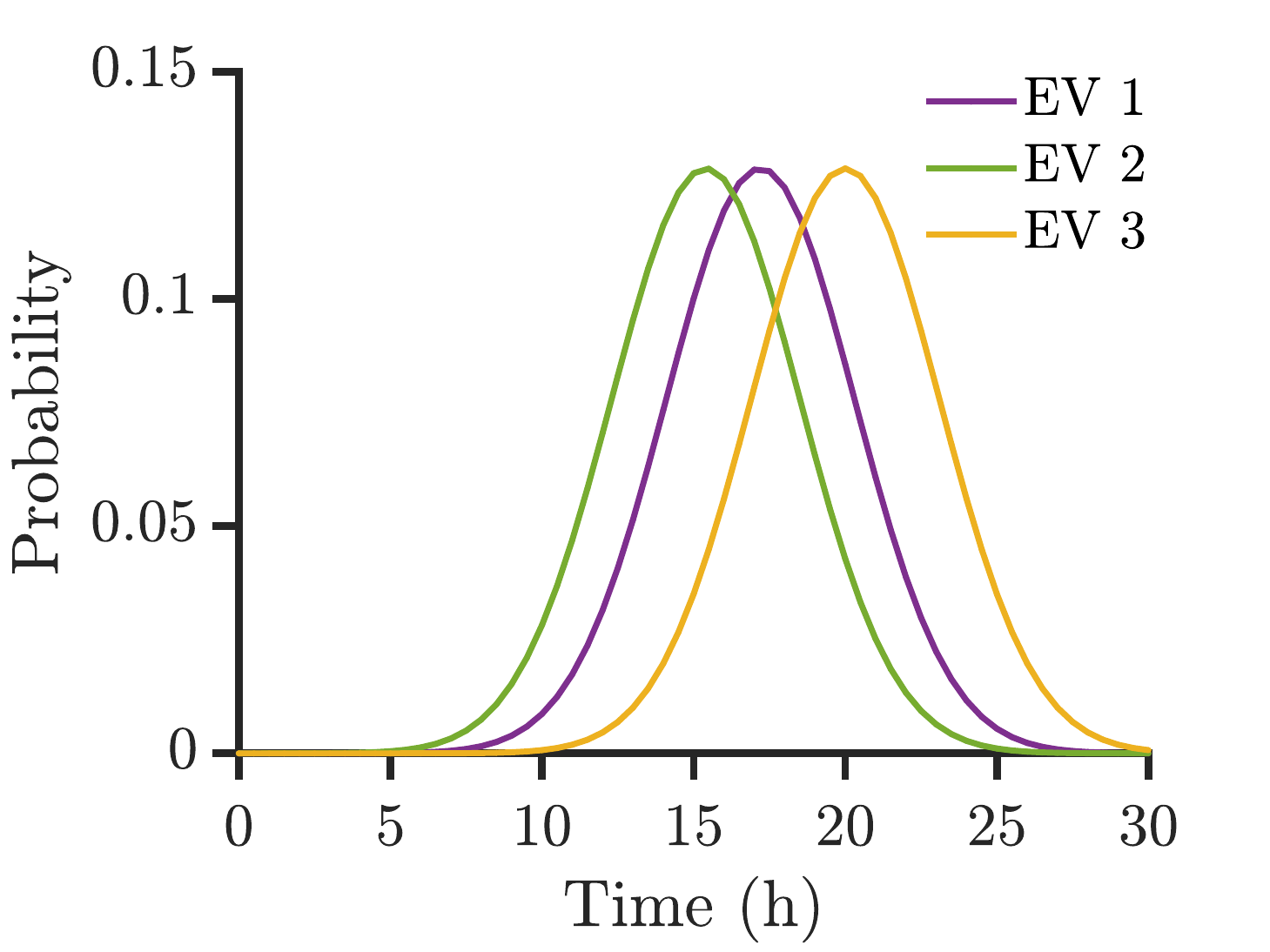}
}
\subfigure[Public departure]{
\includegraphics[width=0.29\linewidth]{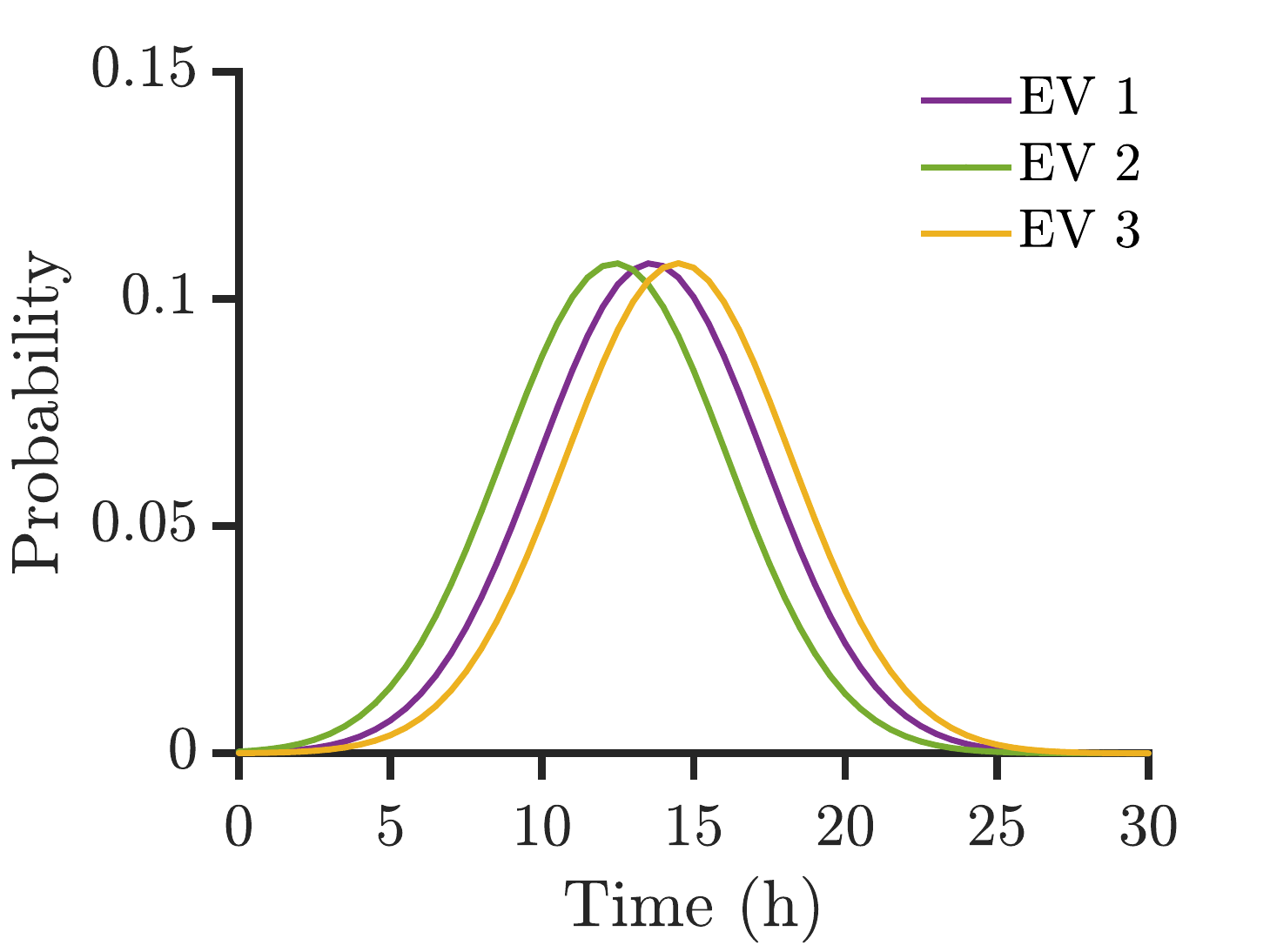}
}
\caption{Statistical distributions of different EV trips}
\label{distribution}
\end{figure}

In order to reflect different users' charging/discharging behaviors, the shape parameter $d_2^i$ which reflects user's expected SoC by the departure time $t_d^i$, is sampled from $[0.85,0.95]$, $[0.85,0.9]$, $[0.9,0.95]$. The anxious time durations are sampled uniformly from $[1, 4]$, $[1, 2]$ and $[2, 4]$. The daily travel plans of EV user 1 follow the statistical analysis of the 2017 National Household Travel Survey (NHTS2017) \cite{nhs2017}. EV user 2 tends to go to work early and return home early as well. EV user 3 prefers to work overtime and always comes home late. Besides, we introduce some distributions related to the arrival and departure times in different places, comprising the home, the office and the public, to depict the different travel plans of drivers. The details of these distributions are shown in Fig. \ref{distribution}, and the time over 24 hours reflects the user has the probability of leaving or getting to someplace on the next day. 

In the SAC setting, each policy network $\pi_{\theta_i}$ contains four serial fully connected hidden layers $\{128,128,128,128\}$ to map the state $s_t^i$. Then two parallel fully connected layers are used to obtain the output $\mu$ and $\sigma$. Meanwhile, RELU is adopted as the activation function to bound the output. Finally the action $a_t^i$ is generated by re-parameterization trick. On the other hand, the current state $s_t^i$ and action $a_t^i$ are concatenated in the input layer of the critic network, and through the hidden layers $\{128,128,128,1\}$ with three RELU functions and one linear rectification function. In the end, the critic network finally ouputs the $q$ value to help the policy evaluation step. The learning rate for actor, critic and temperature parameter $\alpha$ are set as $10^{-3}$, $10^{-2}$ and $10^{-2}$ respectively. The number of training episodes is 250 and the discount factor $\gamma$ for each agent is 0.99. The batchsize is set as 128 and the replay buffer size is $10^5$.



\begin{figure}[htbp!]
\centering
\subfigure[Reward of agent 1]{
\includegraphics[width=0.45\linewidth]{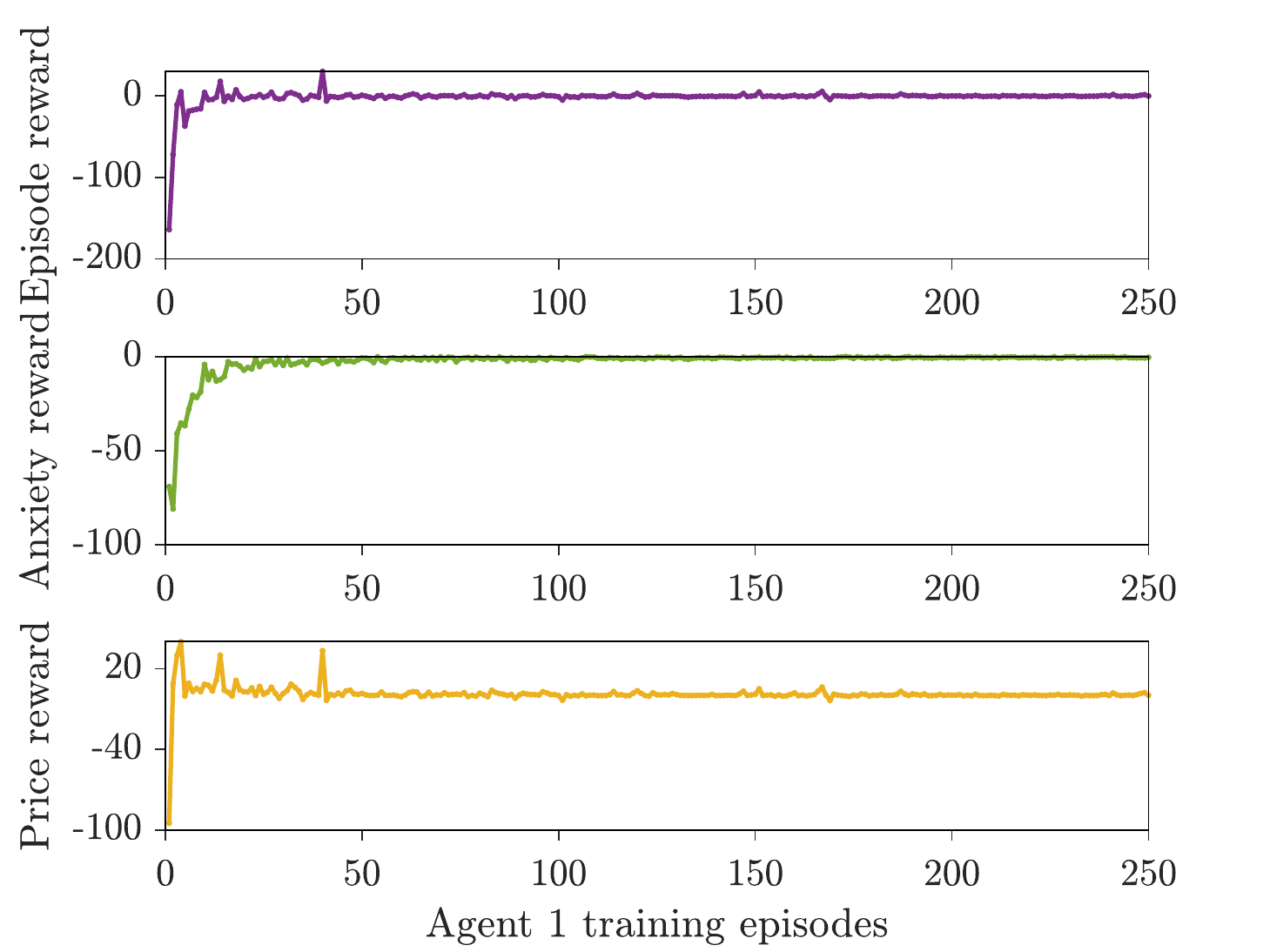}
}
\subfigure[Reward of agent 2]{
\includegraphics[width=0.45\linewidth]{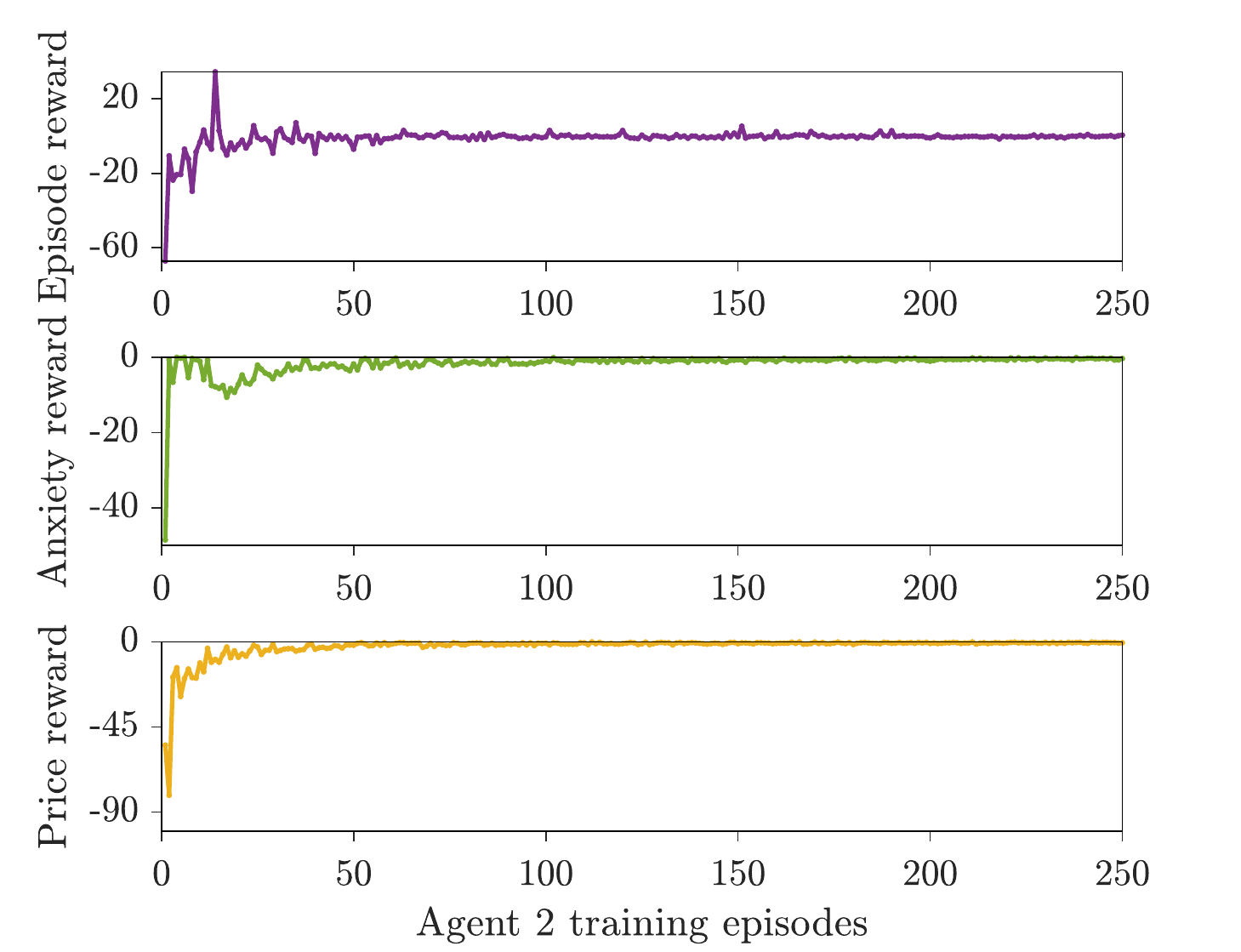}
}
\subfigure[Reward of agent 3]{
\includegraphics[width=0.45\linewidth]{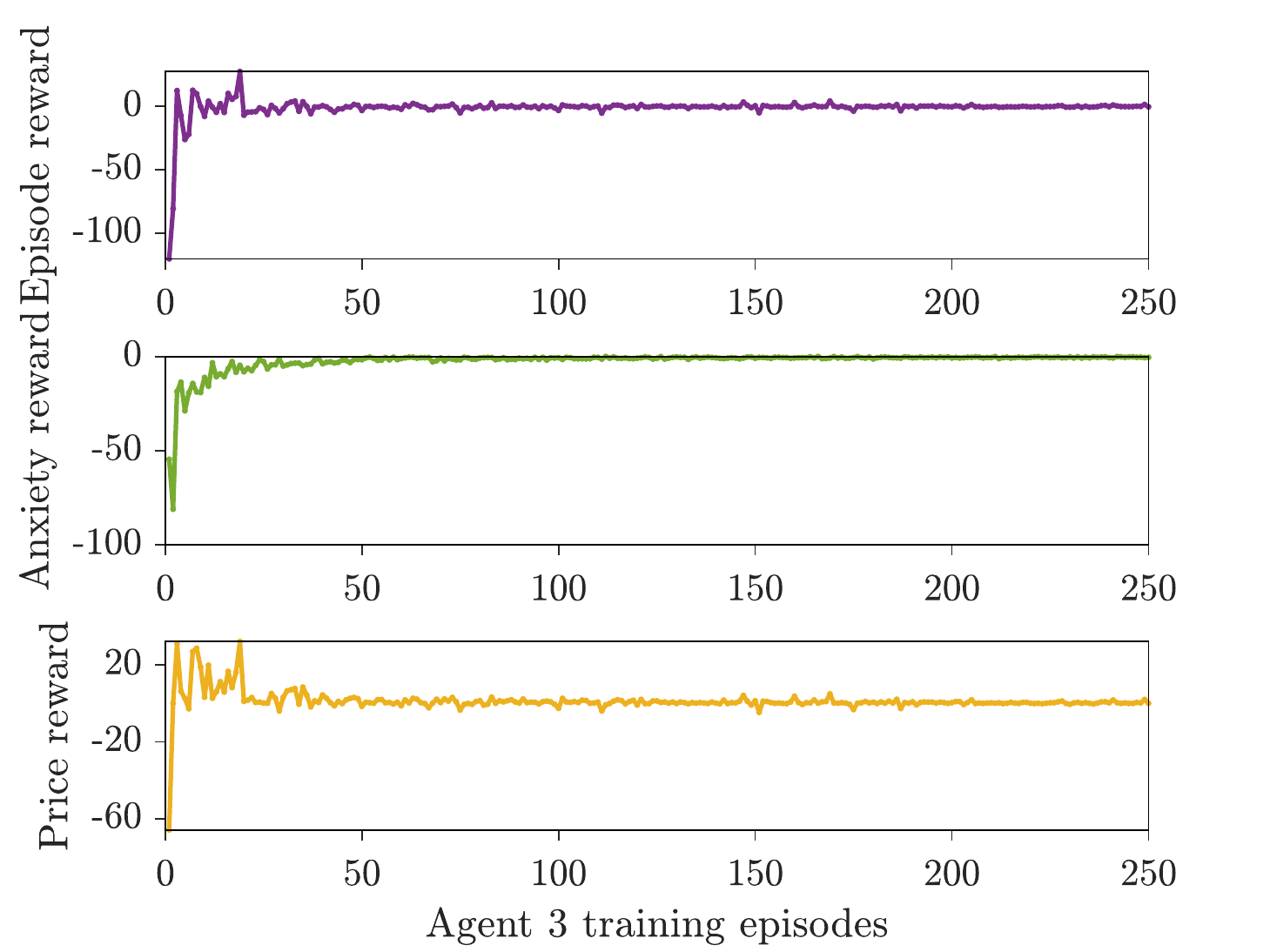}
}
\subfigure[Global average reward]{
\includegraphics[width=0.45\linewidth]{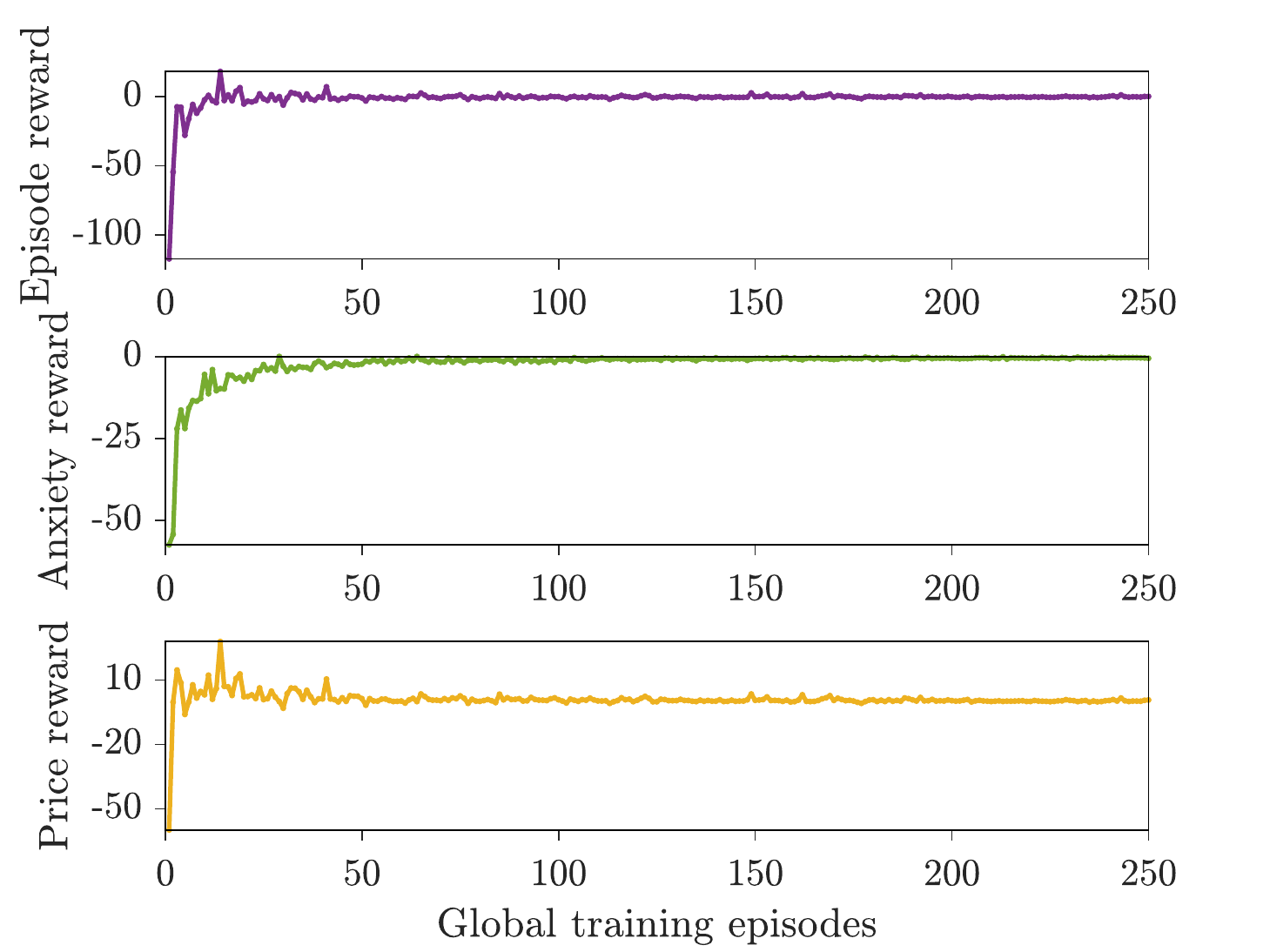}
}
\caption{The local and global training results}
\label{reward}
\end{figure}

The local reward and global average reward during the whole training are shown in Fig. \ref{reward} respectively. Every episode reward can be divided into two parts: the price reward part and the anxiety reward part. The incremental trends of all types of rewards reflect the effect of the proposed approach. As shown in Fig. \ref{reward}(a)$\sim$(d), the curves become more and more stable during the training process, and these rewards all converge finally.

\subsection{Simulation}
We use one week's consecutive trip records which consider three different user charging/discharging behaviors to verify the performance of the proposed approach. Users may choose to travel at different times according to their time schedule. In order to depict the user's dynamic behavior, for a single user, the departure time, arrival time, anxious time and driving time could be different in each trip. In our simulation, drivers depart from home to office in the morning and come home from office in the afternoon during the weekday. On weekends, EV users drive to the public area and EVs are mainly parked in public in the daytime. When EV is in a driving pattern, we assume 5\% of capacity is consumed per hour. When EV is not in use, it will be plugged to the grid by and its charging/discharging rate will be adjusted automatically.

\begin{figure}[htbp!]
\centering
\subfigure[EV 1 SoC]{
\includegraphics[width=0.46\linewidth]{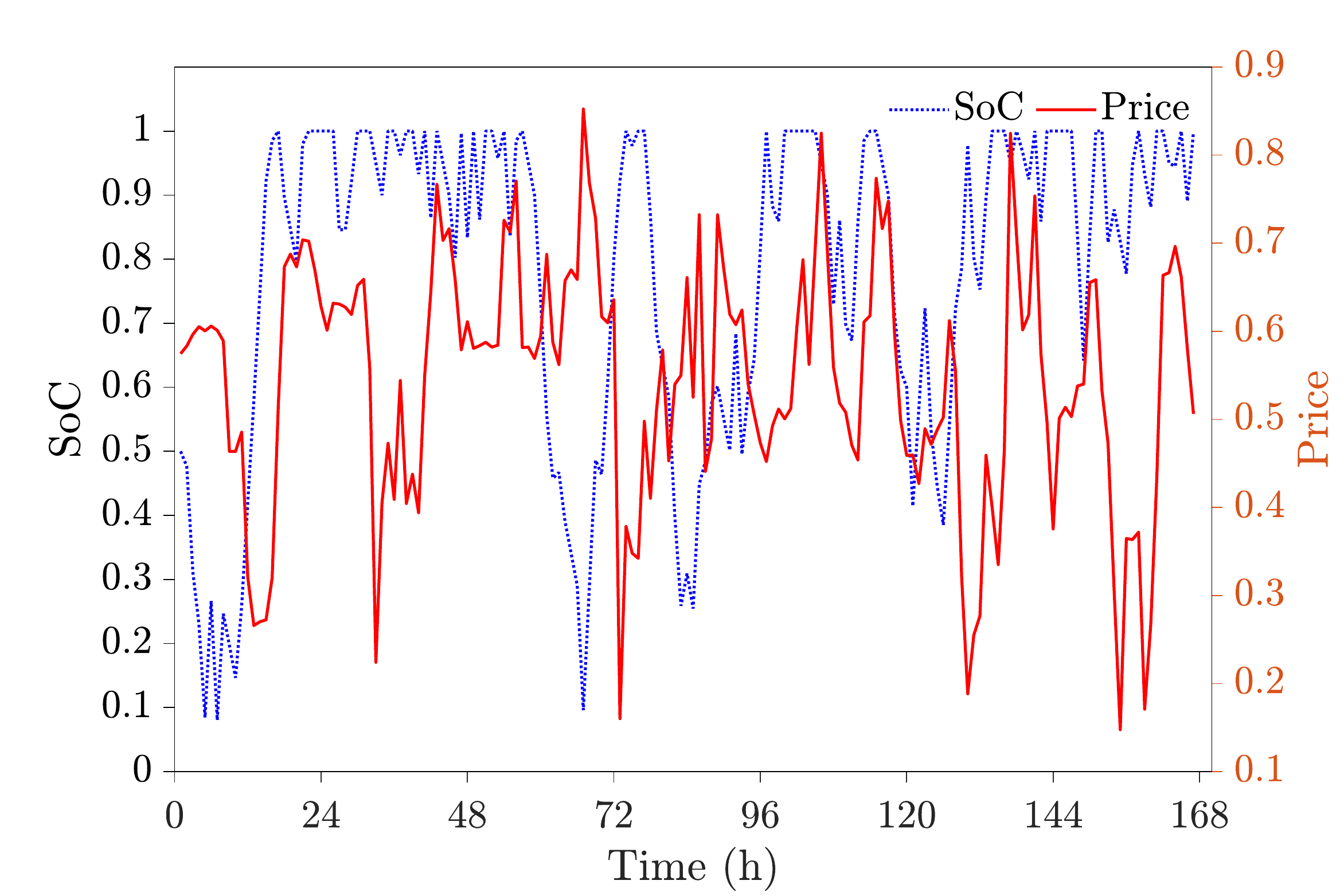}
}
\subfigure[EV 1 charging/discharging power]{
\includegraphics[width=0.46\linewidth]{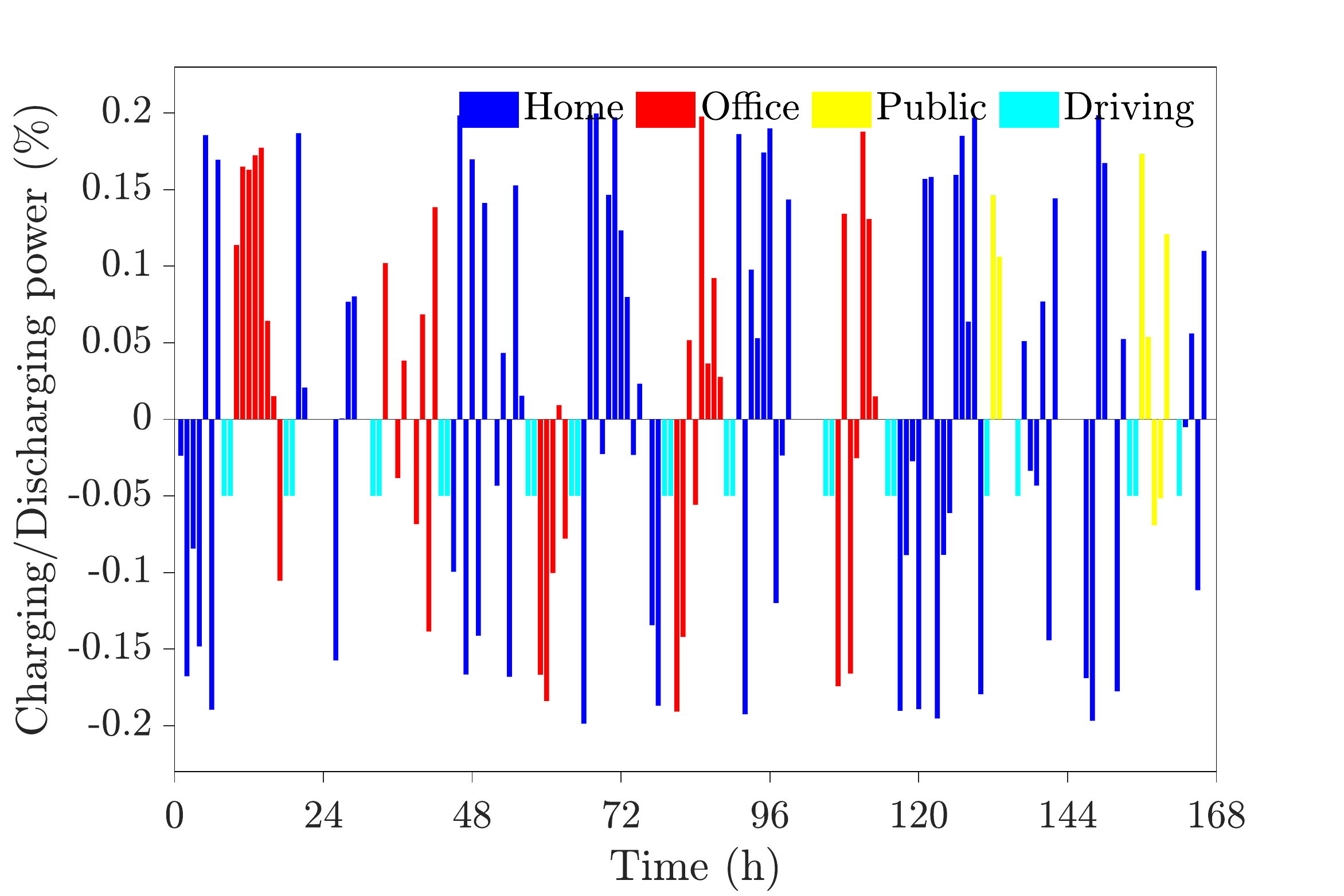}
}
\subfigure[EV 2 SoC]{
\includegraphics[width=0.46\linewidth]{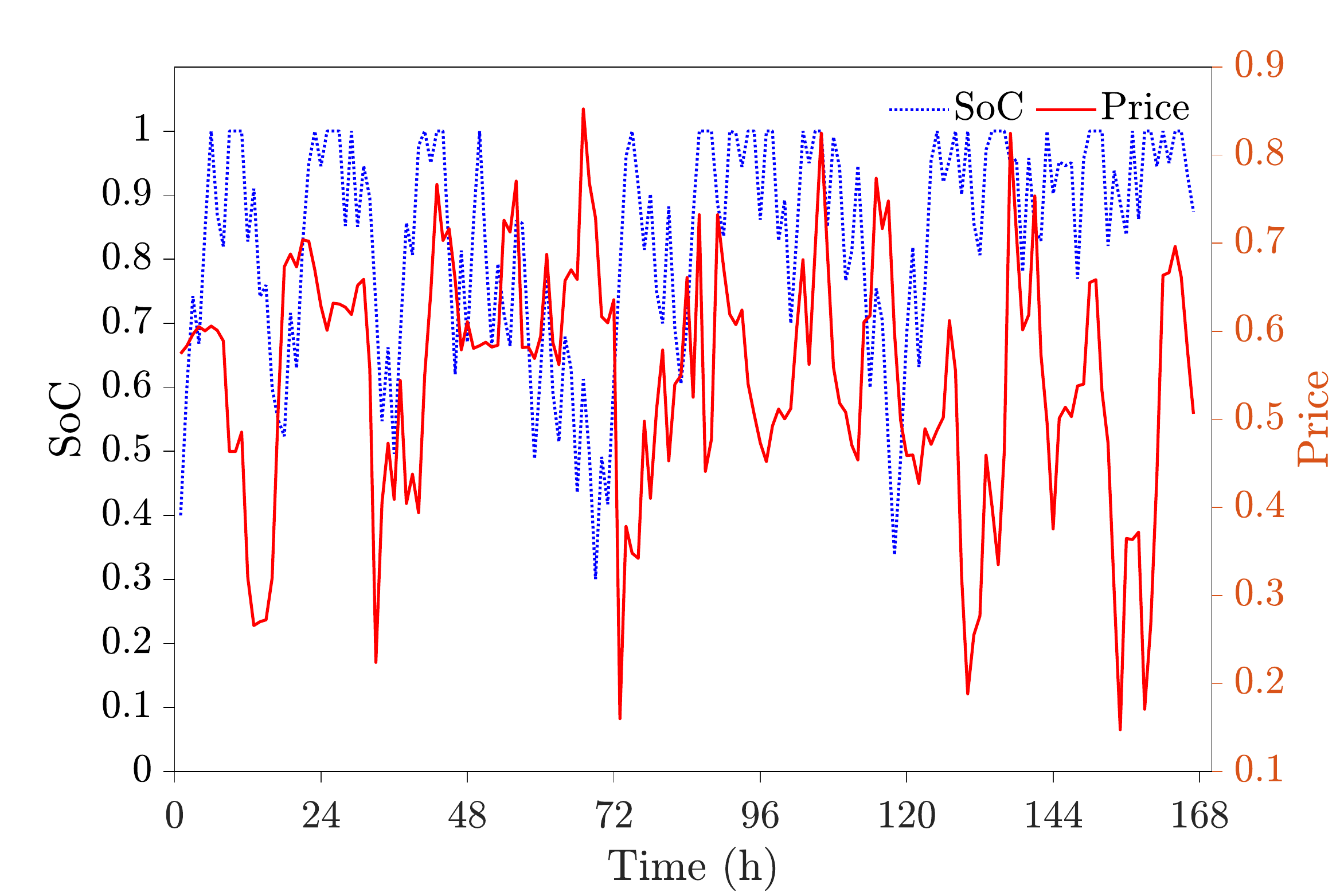}
}
\subfigure[EV 2 charging/discharging power]{
\includegraphics[width=0.46\linewidth]{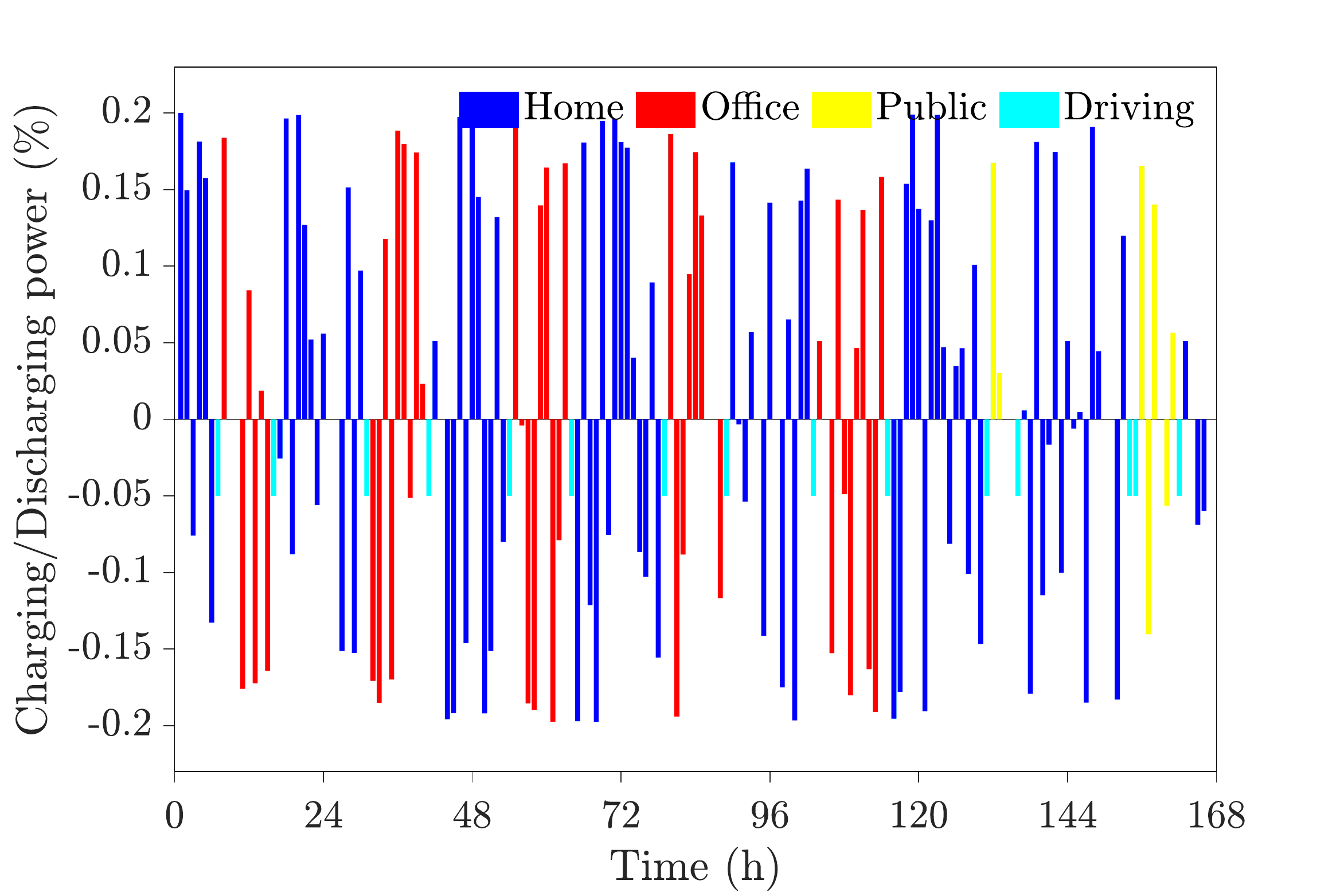}
}
\subfigure[EV 3 SoC]{
\includegraphics[width=0.46\linewidth]{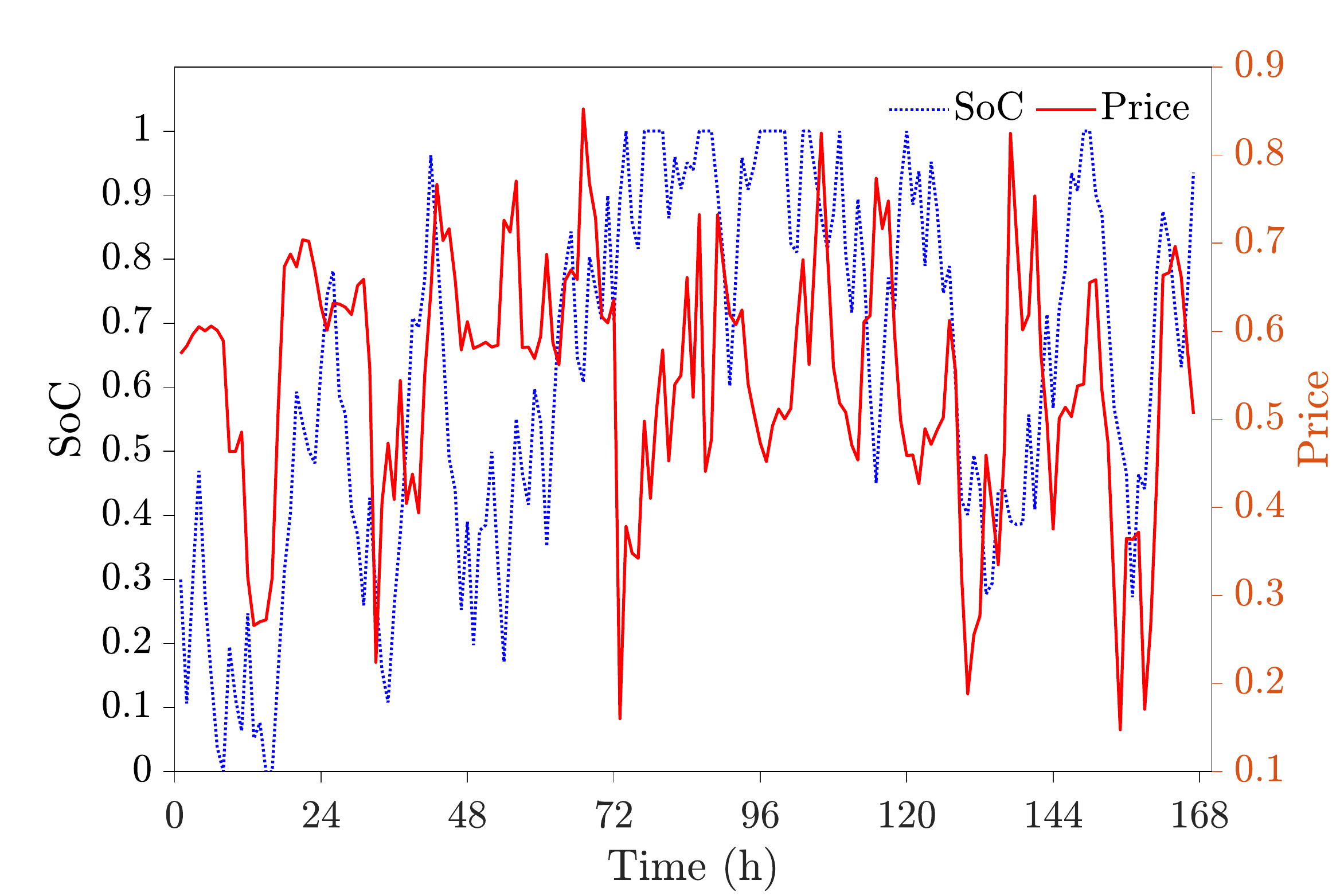}
}
\subfigure[EV 3 charging/discharging power]{
\includegraphics[width=0.46\linewidth]{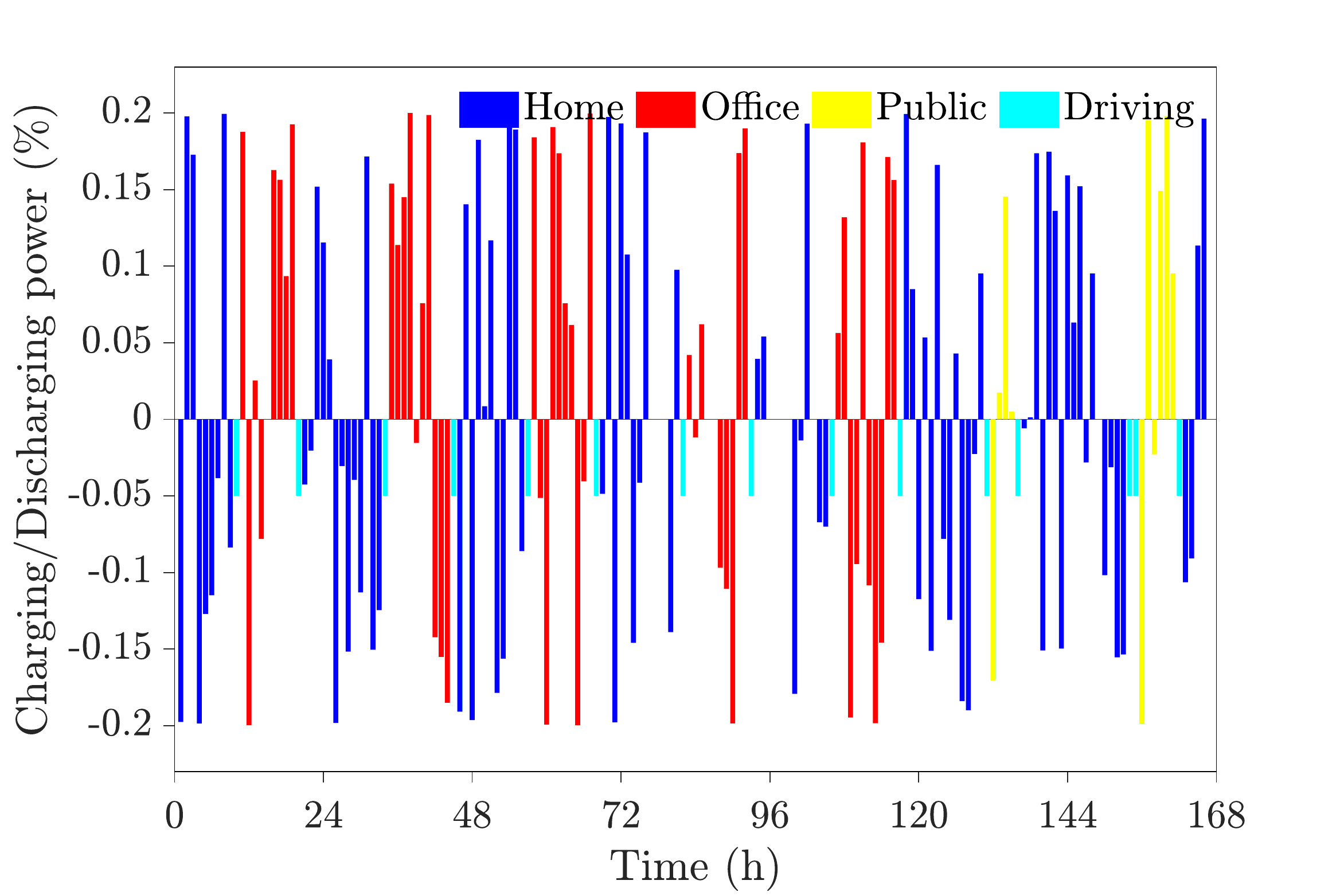}
}
\caption{EV charging/discharging control strategy performance during one week's trip}
\label{simulation}
\end{figure}

The simulation results are shown in Fig. \ref{simulation}(a)$\sim$(f). These charging/discharging actions from different EVs balance well between dynamic electricity prices and their users' travel plans. EVs are usually in the charging mode when the electricity price is low and in the discharging mode when it is high. But if users need to travel during times of the high electricity price, higher charging cost may be required to ensure their travel plans.

\section{Conclusion}
In this paper, a HFRL-based approach was proposed to collaboratively learn a real-time EV charging/discharging control strategy which could perform uniformly well in different kinds of environments. Simulation results showed that the proposed real-time EV charging/discharging control strategy could achieve great performance under the dynamic electricity prices and uncertain users' charging/discharging behaviors.

\section*{Acknowledgement}
\scriptsize
This work was partly supported by the National Nature Science Foundation of China under Grant No. 61601290, and the Shanghai Sailing Program under Grant No. 16YF1407700.
\bibliographystyle{IEEEtran}
\bibliography{refs}

\begin{thebibliography}{10}
\providecommand{\url}[1]{#1}
\csname url@samestyle\endcsname
\providecommand{\newblock}{\relax}
\providecommand{\bibinfo}[2]{#2}
\providecommand{\BIBentrySTDinterwordspacing}{\spaceskip=0pt\relax}
\providecommand{\BIBentryALTinterwordstretchfactor}{4}
\providecommand{\BIBentryALTinterwordspacing}{\spaceskip=\fontdimen2\font plus
\BIBentryALTinterwordstretchfactor\fontdimen3\font minus
  \fontdimen4\font\relax}
\providecommand{\BIBforeignlanguage}[2]{{%
\expandafter\ifx\csname l@#1\endcsname\relax
\typeout{** WARNING: IEEEtran.bst: No hyphenation pattern has been}%
\typeout{** loaded for the language `#1'. Using the pattern for}%
\typeout{** the default language instead.}%
\else
\language=\csname l@#1\endcsname
\fi
#2}}
\providecommand{\BIBdecl}{\relax}
\BIBdecl

\bibitem{chan2007state}
C.~C. Chan, ``The state of the art of electric, hybrid, and fuel cell
  vehicles,'' \emph{Proc. IEEE Proc. IRE}, vol.~95, no.~4, pp. 704--718, 2007.

\bibitem{kempton2005vehicle}
W.~Kempton and J.~Tomi{\'c}, ``Vehicle-to-grid power implementation: From
  stabilizing the grid to supporting large-scale renewable energy,''
  \emph{Journal of power sources}, vol. 144, no.~1, pp. 280--294, 2005.

\bibitem{gan2012optimal}
L.~Gan, U.~Topcu, and S.~H. Low, ``Optimal decentralized protocol for electric
  vehicle charging,'' \emph{IEEE Trans. Power Syst.}, vol.~28, no.~2, pp.
  940--951, 2012.

\bibitem{6102330}
W.~Shi and V.~W. Wong, ``Real-time vehicle-to-grid control algorithm under
  price uncertainty,'' in \emph{2011 IEEE International Conference on Smart
  Grid Communications (SmartGridComm)}, 2011, pp. 261--266.

\bibitem{zhang2020cddpg}
F.~Zhang, Q.~Yang, and D.~An, ``Cddpg: a deep-reinforcement-learning-based
  approach for electric vehicle charging control,'' \emph{IEEE Internet Things
  J.}, vol.~8, no.~5, pp. 3075--3087, 2020.

\bibitem{yan2021deep}
L.~Yan, X.~Chen, J.~Zhou, Y.~Chen, and J.~Wen, ``Deep reinforcement learning
  for continuous electric vehicles charging control with dynamic user
  behaviors,'' \emph{IEEE Trans. Smart Grid}, vol.~12, no.~6, pp. 5124--5134,
  2021.

\bibitem{ortega2014optimal}
M.~A. Ortega-Vazquez, ``Optimal scheduling of electric vehicle charging and
  vehicle-to-grid services at household level including battery degradation and
  price uncertainty,'' \emph{IET Generation, Transmission \& Distribution},
  vol.~8, no.~6, p. 1007, 2014.

\bibitem{wu2017two}
D.~Wu, H.~Zeng, C.~Lu, and B.~Boulet, ``Two-stage energy management for office
  buildings with workplace ev charging and renewable energy,'' \emph{IEEE
  Trans. Transport. Electrific.}, vol.~3, no.~1, pp. 225--237, 2017.

\bibitem{xu2016dynamic}
Y.~Xu, F.~Pan, and L.~Tong, ``Dynamic scheduling for charging electric
  vehicles: A priority rule,'' \emph{IEEE Trans. Autom. Control}, vol.~61,
  no.~12, pp. 4094--4099, 2016.

\bibitem{shi2018model}
Y.~Shi, H.~D. Tuan, A.~V. Savkin, T.~Q. Duong, and H.~V. Poor, ``Model
  predictive control for smart grids with multiple electric-vehicle charging
  stations,'' \emph{IEEE Trans. Smart Grid}, vol.~10, no.~2, pp. 2127--2136,
  2018.

\bibitem{chics2016reinforcement}
A.~Chi{\c{s}}, J.~Lund{\'e}n, and V.~Koivunen, ``Reinforcement learning-based
  plug-in electric vehicle charging with forecasted price,'' \emph{IEEE Trans.
  Veh. Technol}, vol.~66, no.~5, pp. 3674--3684, 2016.

\bibitem{chen2022reinforcement}
X.~Chen, G.~Qu, Y.~Tang, S.~Low, and N.~Li, ``Reinforcement learning for
  selective key applications in power systems: Recent advances and future
  challenges,'' \emph{IEEE Trans. Smart Grid}, 2022.

\bibitem{li2020real}
H.~Li, Z.~Wan, and H.~He, ``Real-time residential demand response,'' \emph{IEEE
  Trans. Smart Grid}, vol.~11, no.~5, pp. 4144--4154, 2020.

\bibitem{mocanu2018line}
E.~Mocanu, D.~C. Mocanu, P.~H. Nguyen, A.~Liotta, M.~E. Webber, M.~Gibescu, and
  J.~G. Slootweg, ``On-line building energy optimization using deep
  reinforcement learning,'' \emph{IEEE Trans. Smart Grid}, vol.~10, no.~4, pp.
  3698--3708, 2018.

\bibitem{9320523}
A.~A. Zishan, M.~M. Haji, and O.~Ardakanian, ``Adaptive congestion control for
  electric vehicle charging in the smart grid,'' \emph{IEEE Transactions on
  Smart Grid}, vol.~12, no.~3, pp. 2439--2449, 2021.

\bibitem{9716749}
L.~Yan, X.~Chen, Y.~Chen, and J.~Wen, ``A cooperative charging control strategy
  for electric vehicles based on multi-agent deep reinforcement learning,''
  \emph{IEEE Trans. Ind. Informat.}, pp. 1--1, 2022.

\bibitem{letaief2021edge}
K.~B. Letaief, Y.~Shi, J.~Lu, and J.~Lu, ``Edge artificial intelligence for 6g:
  Vision, enabling technologies, and applications,'' \emph{IEEE Journal on
  Selected Areas in Communications}, vol.~40, no.~1, pp. 5--36, 2021.

\bibitem{yang2020federated}
K.~Yang, Y.~Shi, Y.~Zhou, Z.~Yang, L.~Fu, and W.~Chen, ``Federated machine
  learning for intelligent iot via reconfigurable intelligent surface,''
  \emph{IEEE Network}, vol.~34, no.~5, pp. 16--22, 2020.

\bibitem{chen2020joint}
M.~Chen, Z.~Yang, W.~Saad, C.~Yin, H.~V. Poor, and S.~Cui, ``A joint learning
  and communications framework for federated learning over wireless networks,''
  \emph{IEEE Trans. Wireless Commun.}, vol.~20, no.~1, pp. 269--283, 2020.

\bibitem{shi2020communication}
Y.~Shi, K.~Yang, T.~Jiang, J.~Zhang, and K.~B. Letaief,
  ``Communication-efficient edge ai: Algorithms and systems,'' \emph{IEEE
  Communications Surveys \& Tutorials}, vol.~22, no.~4, pp. 2167--2191, 2020.

\bibitem{yang2020energy}
Z.~Yang, M.~Chen, W.~Saad, C.~S. Hong, and M.~Shikh-Bahaei, ``Energy efficient
  federated learning over wireless communication networks,'' \emph{IEEE Trans.
  Wireless Commun.}, vol.~20, no.~3, pp. 1935--1949, 2020.

\bibitem{yang2022trustworthy}
Z.~Yang, Y.~Shi, Y.~Zhou, Z.~Wang, and K.~Yang, ``Trustworthy federated
  learning via blockchain,'' \emph{IEEE Internet of Things Journal}, 2022.

\bibitem{yang2022differentially}
Y.~Yang, Y.~Zhou, Y.~Wu, and Y.~Shi, ``Differentially private federated
  learning via reconfigurable intelligent surface,'' \emph{arXiv preprint
  arXiv:2203.17028}, 2022.

\bibitem{wang2021federated}
Z.~Wang, J.~Qiu, Y.~Zhou, Y.~Shi, L.~Fu, W.~Chen, and K.~B. Letaief,
  ``Federated learning via intelligent reflecting surface,'' \emph{IEEE Trans.
  Wireless Commun.}, vol.~21, no.~2, pp. 808--822, 2021.

\bibitem{qi2021federated}
J.~Qi, Q.~Zhou, L.~Lei, and K.~Zheng, ``Federated reinforcement learning:
  techniques, applications, and open challenges,'' \emph{arXiv preprint
  arXiv:2108.11887}, 2021.

\bibitem{zhang2021cooperative}
M.~Zhang, Y.~Jiang, F.-C. Zheng, M.~Bennis, and X.~You, ``Cooperative edge
  caching via federated deep reinforcement learning in fog-rans,'' in
  \emph{2021 IEEE International Conference on Communications Workshops (ICC
  Workshops)}.\hskip 1em plus 0.5em minus 0.4em\relax IEEE, 2021, pp. 1--6.

\bibitem{jin2022federated}
H.~Jin, Y.~Peng, W.~Yang, S.~Wang, and Z.~Zhang, ``Federated reinforcement
  learning with environment heterogeneity,'' in \emph{International Conference
  on Artificial Intelligence and Statistics}.\hskip 1em plus 0.5em minus
  0.4em\relax PMLR, 2022, pp. 18--37.

\bibitem{haarnoja2018soft}
T.~Haarnoja, A.~Zhou, P.~Abbeel, and S.~Levine, ``Soft actor-critic: Off-policy
  maximum entropy deep reinforcement learning with a stochastic actor,'' in
  \emph{International conference on machine learning}.\hskip 1em plus 0.5em
  minus 0.4em\relax PMLR, 2018, pp. 1861--1870.

\bibitem{mcmahan2017communication}
B.~McMahan, E.~Moore, D.~Ramage, S.~Hampson, and B.~A. y~Arcas,
  ``Communication-efficient learning of deep networks from decentralized
  data,'' in \emph{Artificial intelligence and statistics}.\hskip 1em plus
  0.5em minus 0.4em\relax PMLR, 2017, pp. 1273--1282.

\bibitem{alsabbagh2020distributed}
A.~Alsabbagh, B.~Wu, and C.~Ma, ``Distributed electric vehicles charging
  management considering time anxiety and customer behaviors,'' \emph{IEEE
  Trans. Ind. Informat.}, vol.~17, no.~4, pp. 2422--2431, 2020.

\bibitem{oedi_4033}
\BIBentryALTinterwordspacing
G.~Rhodes, ``Electricity market and nrel atb resources,'' 01 2020. [Online].
  Available: \url{https://data.openei.org/submissions/4033}
\BIBentrySTDinterwordspacing

\bibitem{nhs2017}
\BIBentryALTinterwordspacing
F.~H.~A. U.S. Department~of Transportation, ``2017 national household travel
  survey.'' [Online]. Available: \url{http://nhts.ornl.gov}
\BIBentrySTDinterwordspacing

\end{thebibliography}


\end{document}